\documentclass[preprint,12pt,authoryear]{elsarticle}

\usepackage[english]{babel}
\usepackage[utf8x]{inputenc}
\usepackage{amsmath}
\usepackage{amssymb}
\usepackage{soul}
\usepackage{appendix}

\usepackage{amsthm}
\usepackage{graphicx}
\usepackage{multirow}
\usepackage{caption}
\usepackage{subcaption}
\usepackage[table]{xcolor}
\usepackage[colorinlistoftodos]{todonotes}
 \usepackage[margin=1.0in]{geometry}
\usepackage{natbib}
\numberwithin{equation}{section}

\newcommand{\comments}[1]{}

\newcommand{\bvec}[1]{\mathbf{#1}}

\newcommand{\N}[0]{\bvec{n}}
\newcommand{\X}[0]{\bvec{x}}

\newcommand{\dV}[0]{\mathrm{d}V}
\newcommand{\dtet}[0]{\mathrm{d}\Omega_t}
\newcommand{\tetdom}[0]{{\Omega_t}}

\title{An exact general remeshing scheme \\ applied to physically conservative voxelization}

\author[kipac]{Devon Powell}
\ead{dmpowel1@stanford.edu}
\author[kipac]{Tom Abel}
\ead{tabel@stanford.edu}
\address[kipac]{Kavli Institute for Particle Astrophysics and Cosmology, Stanford University, SLAC National Accelerator Laboratory, Menlo Park, CA 94025, USA}

\begin{document}
\begin{abstract}

  We present an exact general remeshing scheme to compute analytic 
  integrals of polynomial functions over the intersections 
  between convex polyhedral cells of old and new meshes.
  In physics applications
  this allows one to ensure global mass, momentum, and energy
  conservation while applying higher-order polynomial interpolation. 
  We elaborate on applications of our
  algorithm arising in the analysis of cosmological N-body data,
  computer graphics, and continuum mechanics problems.

  We focus on the particular case of remeshing tetrahedral cells onto a
  Cartesian grid such that the volume integral of the
  polynomial density function given on the input mesh is guaranteed to equal the
  corresponding integral over the output mesh. We refer to this as
  ``physically conservative voxelization''.

   At the core of our method is an algorithm for intersecting two
  convex polyhedra by successively clipping one against the faces of
  the other. This algorithm is an implementation of the ideas
  presented abstractly by \cite{sugihara1994}, who suggests using the
  planar graph representations of convex polyhedra to ensure
  topological consistency of the output. This makes our implementation
  robust to geometric degeneracy in the input. We employ a simplicial
  decomposition to calculate moment integrals up to quadratic order
  over the resulting intersection domain.

  We also address practical issues arising in a software
  implementation, including numerical stability in geometric
  calculations, management of cancellation errors, and extension to
  two dimensions.  In a comparison to recent work, we show substantial
  performance gains. We provide a C implementation intended to be a
  fast, accurate, and robust tool for geometric calculations on
  polyhedral mesh elements.

\end{abstract}


\maketitle

\section{Introduction}

Several areas of computational physics require one to remesh (also ``remap'' or ``resample'') physical quantities
between meshes made of convex polyhedra in a physically conservative manner. 
By ``physically conservative,'' we mean that
mesh cells from the old and new meshes are overlain, the volumes of intersection between
old and new cells constructed, and the quantity of interest transferred from the old to the new
mesh cells such that the total integral over the output and input are equal.

One instance 
of this is in numerical hydrodynamics, where a highly distorted mesh must be relaxed and remeshed in
order to avoid loss of accuracy. In this context, the subject of this paper is known as a ``direct remap.''
This is of interest in some flavors of Arbitrary Lagrangian-Eulerian (ALE) hydrodynamics (see e.g.
\citealt{Donea2004}) and, more recently, in the ``re-ALE'' scheme pioneered by \cite{loubere2010}.
The precise problem of intersecting arbitrary polyhedra for direct remapping of hydrodynamical
quantities is attacked by \cite{grandy1999}, who gives a description of a first-order
scheme for polyhedral grids, as well as a thorough review of the topic. \cite{dukowicz1987}
and \cite{dukowicz1991} present algorithms for the same problem, including higher-order
interpolation during the remap step.

Interface reconstruction for
multiphase flows (specifically, piecewise-linear interface reconstruction, or PLIC) relies on
calculating the volume of a mesh cell that has been truncated against the interface plane.
This is akin to the
intersection of two polyhedra, though in this case the problem is restricted to intersecting a
polyhedron with another plane.
\cite{hirt1981}
describe the so-called ``volume-of-fluid'' (VOF) methods, which enforce volume conservation of
material in a grid cell during this clipping operation.  \cite{renardy2001} provide a good
overview of the basic concepts involved. \cite{lopez2008} give a Fortran toolkit of the
necessary operations for VOF, against which we give a direct comparison in the results section.

A physically conservative remesh is also useful for visualization purposes. 
Specifically, we refer to ``voxelization'' (also ``scan-line conversion'' or ``rasterization''),
a geometric operation in which polyhedra are mapped onto a 3D Cartesian
lattice of cubical grid cells (``voxels''). 
The current state-of-the-art in voxelization as a computer graphics application is given by 
\cite{duff1989}, who generates anti-aliased images by computing exact
convolution integrals for separable polynomial filters.  More recently, \cite{auzinger2012} and
\cite{auzinger2013} compute exact convolution integrals for non-separable (spherically-symmetric)
polynomial filters. \cite{catmull1978} describes area-sampling in 2D, which is equivalent to convolving
the continuous image with the pixel shape. 
Again, the problem of intersecting convex polyhedra arises, as voxelization on a conceptual level is simply
the computation of integrals over the intersection volumes between cubical grid cells and input polyhedra.

The particular application for which we developed the method presented here is the exact mass-conservative
voxelization of tetrahedra for the simulation and analysis of cosmological $N$-body systems using
the method of \cite{AHK2012}. This approach to the N-body problem treats dark matter
particles as tracers, and interprets the mass as being interpolated between the tracers in tetrahedral
mass elements. This approach has the advantage of giving a well-defined density field everywhere in space,
eliminating the need to consider particle discreteness effects. It has since been explored
in more detail by \cite{angulo2014}, who create smooth maps of the gravitational lensing
potential around dark matter halos, \cite{hahn2013}, who show that this method eliminates
artificial clumping in N-body simulations, and \cite{hahn2014}, who look at statistics of cosmic
velocity fields. \cite{Kaehler2012} use voxelization in a
visualization context to produce stunning and informative renderings of cosmic structures.

\begin{figure}[h!]
\centering
\begin{subfigure}[b]{0.32\textwidth}
	\includegraphics[width=\textwidth]{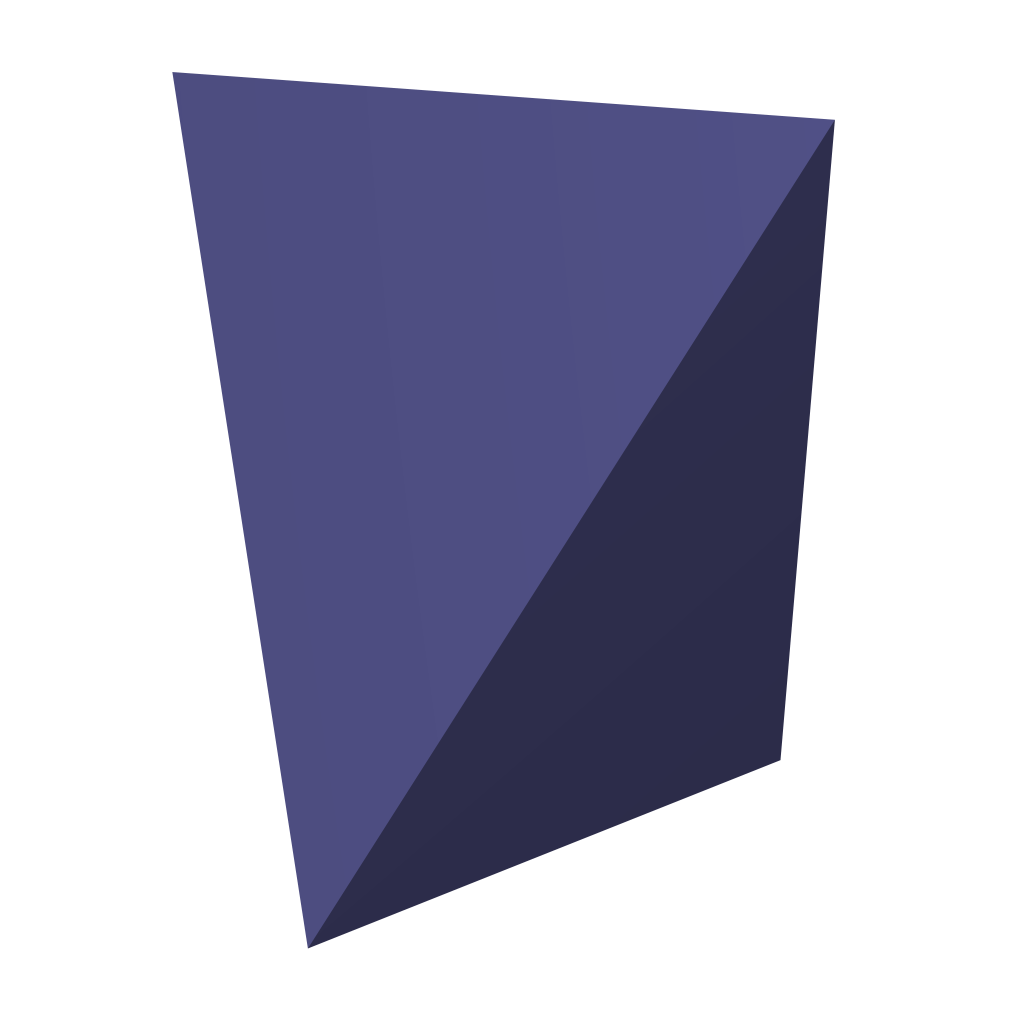}
\end{subfigure}
\begin{subfigure}[b]{0.32\textwidth}
	\includegraphics[width=\textwidth]{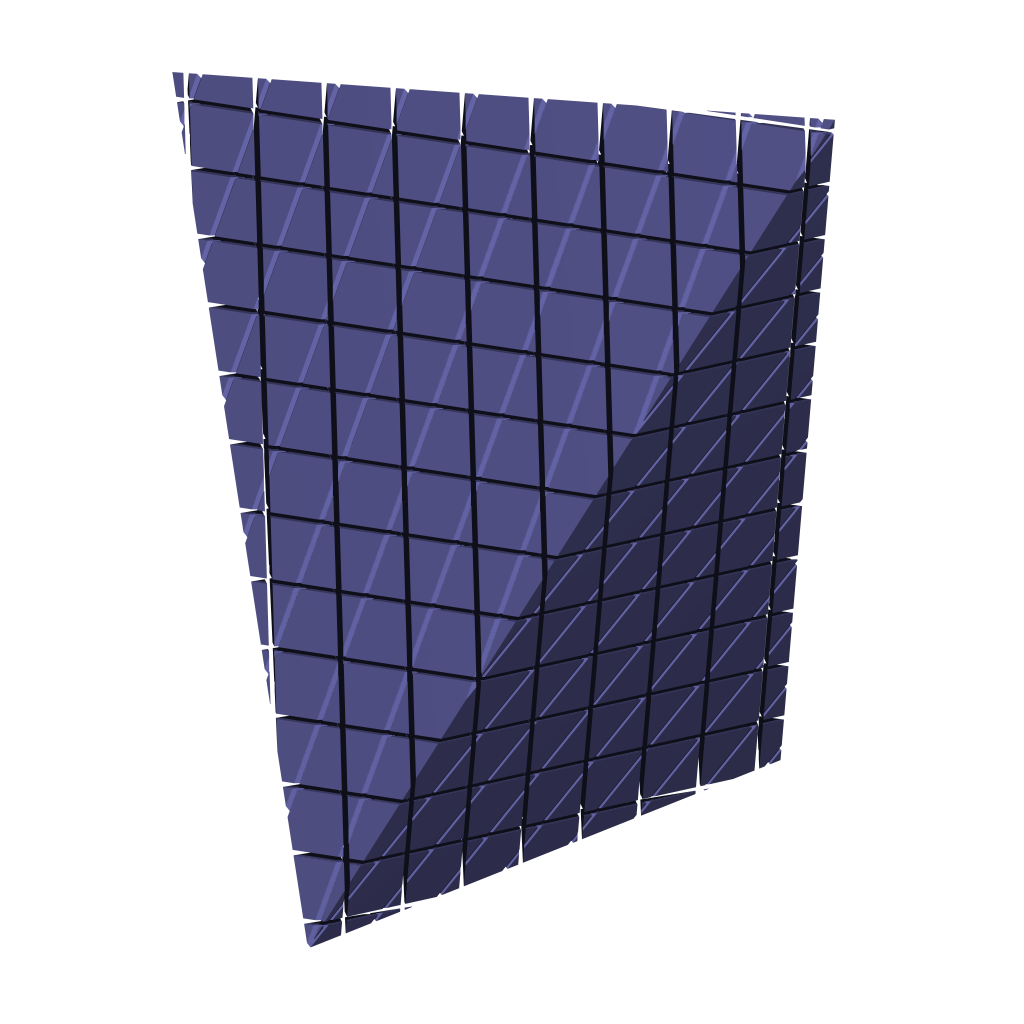}
\end{subfigure}
\begin{subfigure}[b]{0.32\textwidth}
	\includegraphics[width=\textwidth]{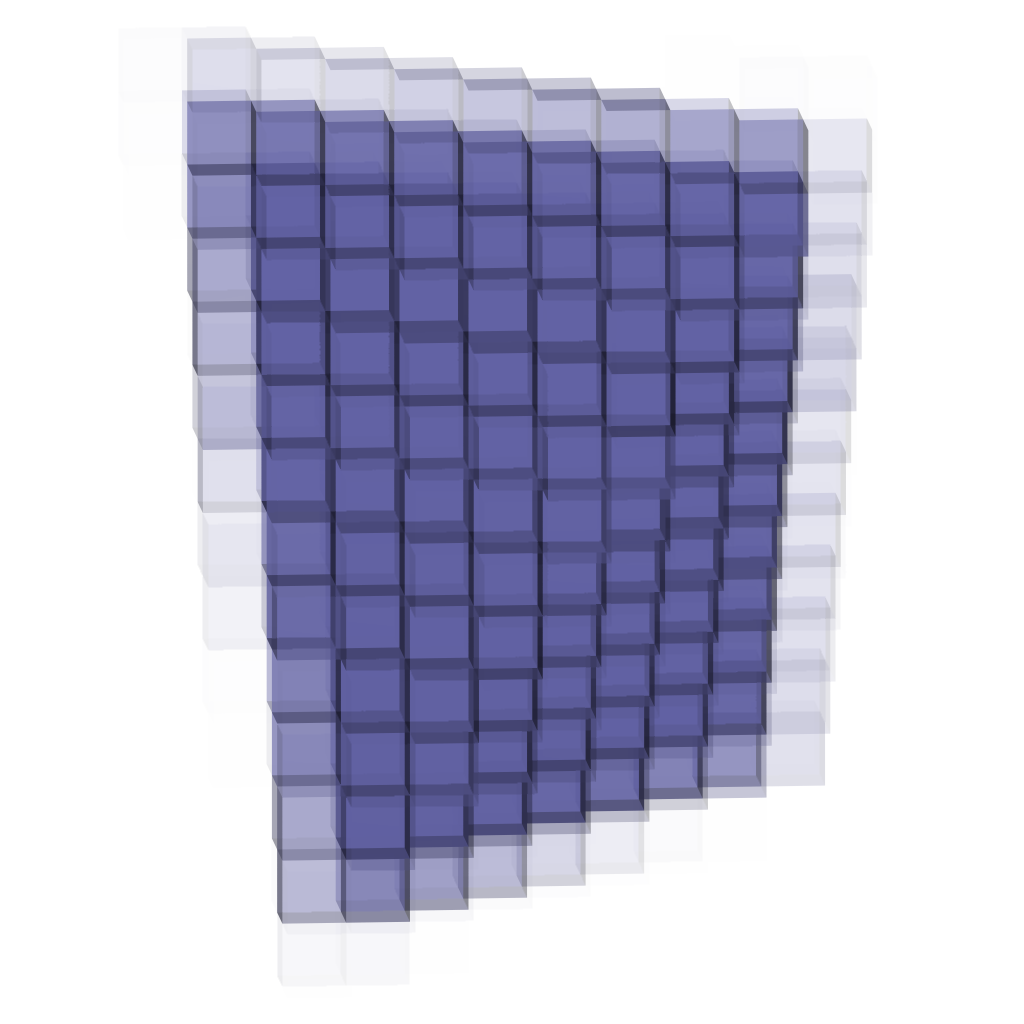}
\end{subfigure}
\caption{An illustration of physically conservative voxelization. Left: An
input tetrahedron. Middle: The input tetrahedron, split between underlying cubical grid cells
(voxels) to form a set of
non-overlapping integration domains, each of which is the intersection of a cube and the tetrahedron. Right: the voxelized
tetrahedron. The total volume integral is conserved to high precision between the left and right figures.}
\label{fig:voxelization3d}
\end{figure}

\begin{figure}[h!]
\centering
\includegraphics[width=1.0\textwidth]{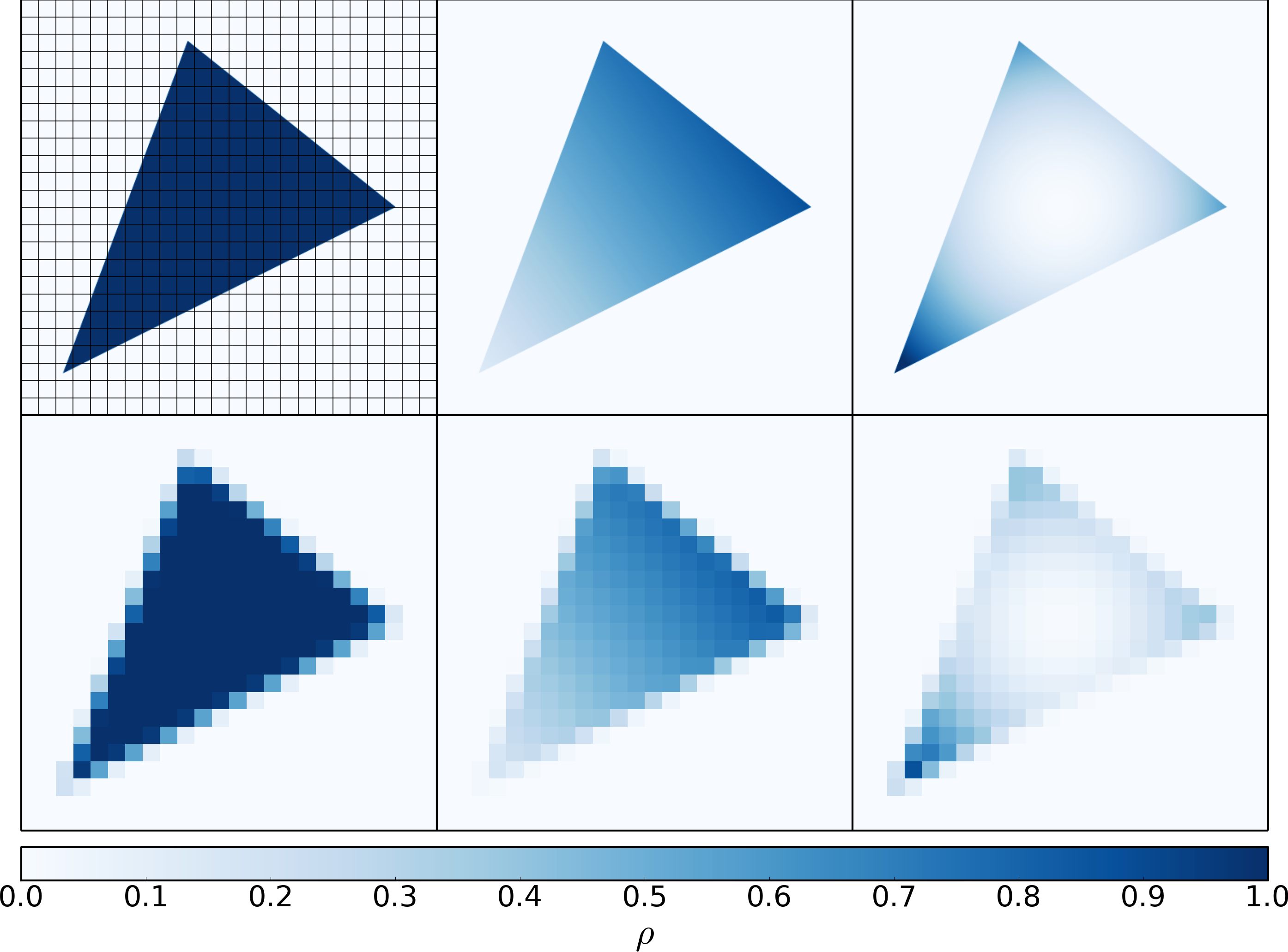}
\caption{Another illustration of physically conservative voxelization. Here we show slices through a voxelized
triangular prism with constant (left), linear (middle) and quadratic (right) density fields defined
over the domain. Top row: The input, with continuous polynomial density. The top-left panel
also shows the grid to which we are voxelizing. Bottom row: The voxelized output. Mass is conserved
to high precision between the top and bottom rows.}
\label{fig:raster}
\end{figure}

We refer in this paper to the specific case of ``physically conservative voxelization'' of tetrahedra,
in which a scalar density defined across the input tetrahedron is integrated over each domain
formed by the intersection of the tetrahedron with each grid cell that intersects the
tetrahedron (see Figures \ref{fig:voxelization3d} and \ref{fig:raster} for illustrations).
Hence, the sum over each voxel in the output should exactly equal the total integral over the input.

Thus, this paper describes a specific application of a physically conservative remesh, while
remaining cognizant of the fact
that the concepts presented here form a general conservative remesh scheme applicable to any of the
aforementioned problems.  Our goal is to present a unified approach for intersecting two convex polyhedra in a
geometrically robust way, and for accurately computing the integral of a polynomial function over the
resulting intersection domain.

As we discuss in Section \ref{sec:algorithm}, there are two basic operations that form the core of
our algorithm: a clipping operation, in which a cube is repeatedly truncated against the faces of a
tetrahedron (this is equivalent to intersecting two convex polyhedra),
and a reduction operation, in which we integrate a polynomial density over the
the convex polyhedron resulting from the clipping operation.

The integration of polynomial fields over arbitrary polyhedral domains has been well-studied in the
literature. One approach is to reduce the dimensionality of a volume integral using the divergence theorem.
\cite{dukowicz1987}, \cite{liggett1988}, \cite{mirtich1996}, and \cite{margolin2003} all use the divergence
theorem to reduce such volume integrals to line integrals. In particular, the two-dimensional case in
Cartesian coordinates is well known (\citealt{Stone:1986}), including formulae for moments over the
polygons (\citealt{Bockman:1989}). A second approach to this problem is to decompose the domain into simplices (tetrahedra in 3D,
  triangles in 2D) and carry out the
integration over each simplex separately using existing formulae derived using barycentric
coordinates (see Section \ref{sec:reduction}). This method is well-established in implementations of the finite element method (FEM),
where mesh elements are often simplices.  Most recently, \cite{deloera2011} present software for the
exact integration of polynomials over convex polyhedral domains using simplicial decomposition.
\cite{liu1998} present a similar algorithm.

The clipping operation is more subtle. The classic method for clipping a polyhedron against
a plane is Sutherland-Hodgman clipping (\citealt{sutherland1974}), which simply tests vertices
against the clip plane and excludes those vertices lying on the ``wrong'' side.  This method
requires some sort of lookup table for reconstructing the boundary connectivity of the resulting
polyhedron, as implemented by \cite{stephenson1975}. However, this
method is not geometrically robust, meaning that it can admit geometrical inconsistencies
if vertices lie on the clip plane to within roundoff error. In this case, vertices may be duplicated
or omitted, giving an invalid representation for the
output polyhedron. This is catastrophic for, say,
the integration process, which requires complete and self-consistent geometrical information.

Because ours is a computational physics application, geometrical robustness to the input data is a
necessity; we require accurate and conservative output for \emph{all} possible input cases.
Previous work in computational physics (e.g. \citealt{dukowicz1991}, \citealt{grandy1999}) has dealt with
this issue by using auxiliary algorithms for detecting and artificially removing such geometrical
ambiguities, or by dealing with them on a case-by-case basis.  We instead propose to handle
geometric degeneracies in a cleaner way, by implementing an algorithm that is naturally immune to
them, and hence manifestly robust.

The question of how to design a geometrically robust method for intersecting polyhedra (intersecting
two convex polyhedra is equivalent to repeatedly clipping one against the faces of the other) is
addressed in detail by \cite{stewart1994}, who gives a thorough review of the literature.
The author divides ways of achieving geometric
robustness into three main classes. The first is exact arithmetic, meaning that the input polyhedron is
represented exactly (e.g. integer or rational coordinates), removing the possibility of geometric
ambiguity from subsequent tests. Either some form of
high-precision arithmetic is used (e.g. \citealt{sugihara1990}), or the vertices/edges of an input polyhedron are perturbed
in such a way that a finite-precision algorithm can never encounter a geometrically ambiguous
decision (e.g. \citealt{milenkovic1988}). The second is the representation and model approach developed
by \cite{hoffman1988},
which makes geometric decisions guaranteeing that the mapping from input to output
representations always corresponds to valid input and output models.
The final class are the topological consistency methods.
They work by guaranteeing that the output is always valid in a topological sense.
Such methods are not in general provably robust, but empirical tests have shown that they indeed are.
\cite{sugihara1989} achieve this by eliminating redundant numerical tests.
Other examples include \cite{karasick1989}, who employs rules on how
geometric intersections are allowed to occur, and \cite{bruderlin1991}, who checks nearby features that may be merged
for whether they result in a valid polyhedron.

Our choice of which of these aforementioned algorithms to use for our application is informed by the
fact that we want to perform both the clipping and reduction operations on the same polyhedral
representation, so as to avoid the extra computation needed in changing representations.

We choose to represent convex polyhedra using their planar graphs and perform the clipping
operation in a way that preserves the topological validity of the output graph, as suggested by
\cite{sugihara1994}. This is a member of the topological consistency methods, and it ensures that
our method is robust by making geometric decisions combinatorially, using numerical
comparisons only as a guide. This representation for convex polyhedra lends itself naturally to a simplicial
decomposition approach for the integration step, which can be accomplished easily by traversing the graph.

The layout of this paper is as follows. In Section \ref{sec:motivation}, we discuss the original
application for this work (the analysis and simulation of dark matter in cosmology), as well as
elaborating on other potential applications in computer graphics and hydrodynamics. Section \ref{sec:algorithm}
presents in detail the main concepts in the voxelization algorithm, while Section \ref{sec:practical}
discusses some subtleties arising in a practical implementation. Finally, we present results
concerning accuracy, robustness, and performance of our C implementation, comparing to
previous work, in Section \ref{sec:results}.

\section{Motivation} \label{sec:motivation}

\subsection{Cosmological N-body data}

\begin{figure}[h!]
\centering
\includegraphics[width=1.0\textwidth]{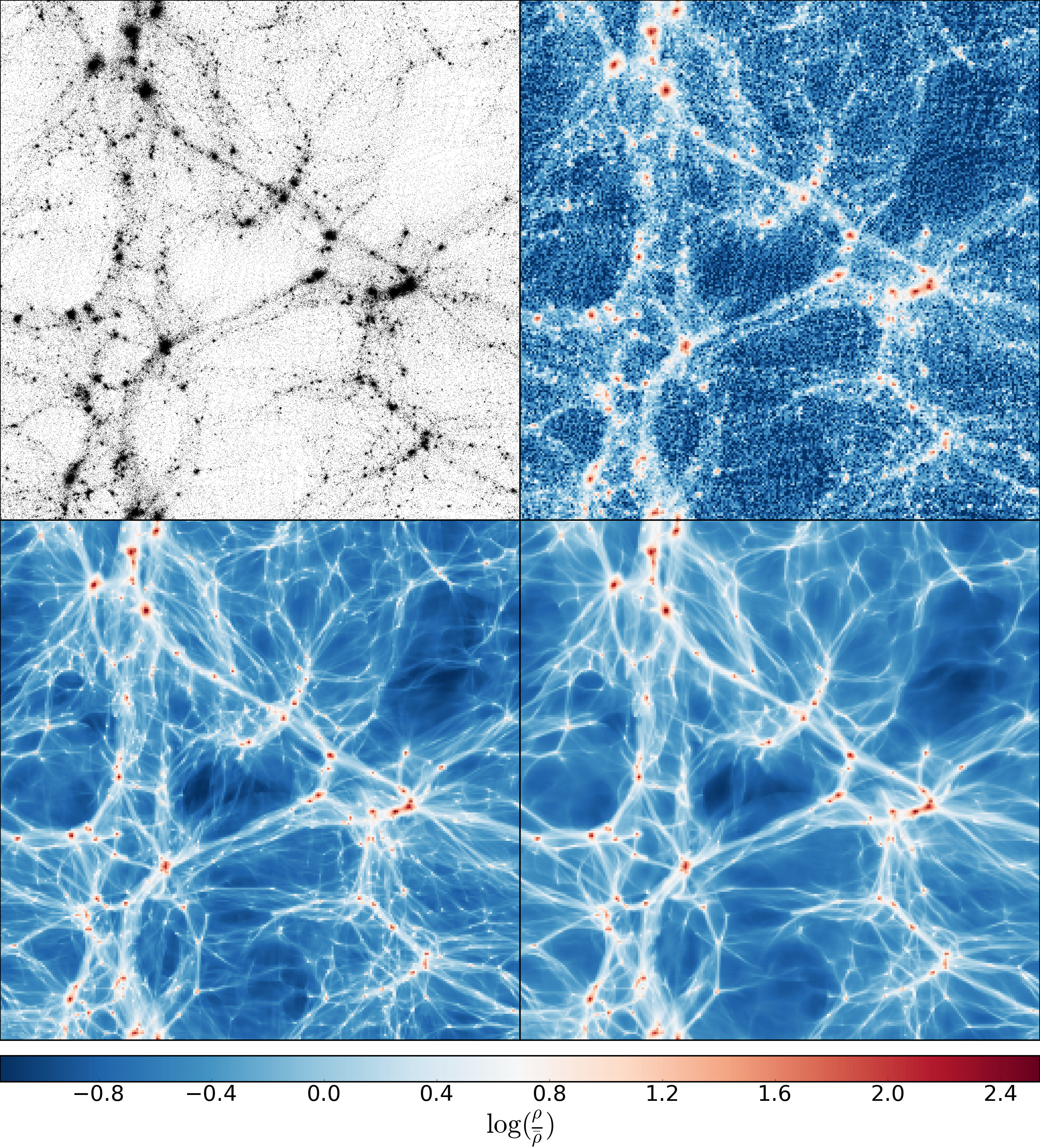}
\caption{Visual comparison of methods for depositing dark matter mass onto a regular grid. Top left: Scatter
plot of particle locations.  Top right: Cloud-in-cell (CIC) deposit. Bottom left: Conservative voxelization, with
piecewise-constant density. Bottom right: Conservative voxelization, with piecewise-linear density.
The bottom two panels use the method of \cite{AHK2012} to properly capture the phase-space structure
of the dark matter, using the physically conservative voxelization scheme presented here to conserve mass to
machine precision. This is the motivation for the present work.}
\label{fig:dm}
\end{figure}

The application for which we developed the method presented here is the physically conservative
voxelization of tetrahedra for the simulation and analysis of cosmological $N$-body systems using the
approach of \cite{AHK2012}. (See Figure \ref{fig:dm}.)

Dark matter, like any continuous system in physics, can be represented by a probability distribution
function (PDF) in phase space.  The computational solution of this system requires the
discretization of this continuous PDF, something traditionally done using Dirac-$\delta$-like particles. As
such, the results of commonly used N-body codes such as Gadget2 (\citealt{springel2005}), HACC
(\citealt{habib2012}), 2HOT (\citealt{warren2013}), Enzo (\citealt{enzo2014}), Ramses
(\citealt{teyssier2002}), NyX (\citealt{almgren2013}), and ART (\citealt{kravtsov1997})
are given in a particle description.  While these codes
differ in the ways they compute gravitational forces, decompose the domain, apply force-softening to
particles, etc., they are all fundamentally N-body codes that interpret dark matter mass as being
concentrated at point locations.

This can be problematic when it comes to the analysis of such N-body data.  In many applications
(e.g. solution of the Poisson equation, identification of cosmic voids, visualization), we desire
the density field to be represented continuously, so that there is no ambiguity in the local density
of a particle distribution.  Simply binning particles into their nearest grid cell (known as
cloud-in-cell, or CIC; see e.g. \citealt{hockney1988}) is a ubiquitous technique for doing so, and is
traditionally used in the force computation step for particle mesh (PM) codes. Voronoi tessellation
around particles has also seen some success (e.g., \citealt{neyrinck2008}). However, both of these methods
are subject to Poisson counting uncertainty due to their particle nature.

\cite{AHK2012} study N-body (dark matter) simulation data by representing the cold phase-space
distribution of dark matter as a three-dimensional “sheet” tessellated into simplices. When modeling a
cold fluid in phase space, one only needs to represent a three-dimensional manifold moving in the
six-dimensional phase space. The tetrahedral tessellation is thus a piecewise-linear approximation
to this smooth Lagrangian three-manifold embedded in $\mathbb{R}^6$.  Rather than carrying mass themselves, 
the particle positions serve merely as tracers of the underlying phase-space distribution; the mass itself is
interpolated between tracer particles with neighboring Lagrangian coordinates. 
Such an approach unambiguously gives the density and velocity
of the distribution everywhere in configuration space, as opposed to traditional particle-based
schemes, which are subject to sampling noise.  When the tracer particles move on characteristics and the
mass inside the volumes they span can be assumed invariant, the full microphysical phase-space
structure is captured by such a three-dimensional sheet.

This scheme has the advantage of giving a well-defined density field everywhere in space,
eliminating the need to consider particle discreteness effects. The method has since been explored
in more detail by \cite{angulo2014}, who create smooth maps of the gravitational lensing
potential around dark matter halos, \cite{hahn2013}, who show that this method eliminates
artificial clumping in N-body simulations, and \cite{hahn2014}, who look at statistics of cosmic
velocity fields.
\cite{Kaehler2012} employ this method in a scientific visualization context, exploring its
advantages over a variety of other techniques and discussing some key ways to
frame the problem so that primitives of typical computer graphics
hardware can be exploited optimally.

This description of dark matter as a collection of mass-carrying tetrahedra is very useful.
However, this density field is still represented in Lagrangian
space; we require a way to project tetrahedral mass elements onto a Cartesian grid in 3D
configuration space.

\cite{angulo2014} recursively split tetrahedra until each one is
smaller than a grid cell, then deposits the mass to the nearest cell.
This conserves mass, but is relatively slow compared to the method
described here, and introduces some small-scale noise.
\cite{hahn2013} use a CIC deposit, the mass-conserving particle-based
method mentioned above, to generate the density field used in solving
the Poisson equation.  This method works well for simulating
gravitational forces between particles, but is unsuitable for
visualization or for more advanced simulation methods such as
\cite{hahn2015}.  \cite{hahn2014} use a multisampling approach, in
which the mass distribution is sampled several times within each grid
cell, and the results averaged.  This method works very well for
visualization and some analysis purposes. However, for certain other
applications, including solving for gravitational forces using the
Poisson equations, we need the total mass to be conserved. 

So, we desire the total mass contained in a grid cell to exactly equal the integral of the input
density field over the cell, while avoiding aliasing noise, and to do it quickly and accurately.
The problem reduces to finding the
integral of a polynomial density field over each domain resulting from the intersection of the input
tetrahedron with the cubical grid cells. In other words, we require a physically conservative voxelization
scheme for depositing tetrahedral mass elements to a grid.

\subsection{Computer graphics and visualization}

As discussed in the introduction, another area for which this work may be 
useful is that of voxelization for computer graphics. This is in the context of computing exact
convolution integrals of polynomial filters over cubical cells. In one way or
another, \cite{catmull1978},  \cite{duff1989}, 
\cite{auzinger2012}, and \cite{auzinger2013} compute such convolution integrals
to achieve perfect anti-aliasing in the output images.
Our method extends the area-sampling approach of \cite{catmull1978} to ``volume-sampling''
in 3D. 

\cite{hasselgren2005}, \cite{zhang2007}, and \cite{pantaleoni2011} describe GPU implementations of
``conservative'' voxelization. In their context, ``conservative'' means that each voxel that
intersects a polyhedron is correctly identified; however,  there is no guarantee that the total
volume integral is conserved.   Our method could be efficiently combined with these
hardware-accelerated collision-finding algorithms to exactly enforce local conservation of physical
quantities. This is exciting for scientific visualization applications such as that of
\cite{Kaehler2012}.

\subsection{Hydrodynamics}

Generally speaking, numerical hydrodynamics schemes are
derived from conservations laws (continuity equations) for mass, momentum, and energy in their
integral forms (see \citealt{jameson1981}, for example). \cite{Hughes1981} and \cite{Donea2004}
describe Arbitrary Lagrangian-Eulerian (ALE) schemes on a moving
mesh. A major component in ALE schemes is the remesh, which moves the mesh to some desired updated
configuration and reapportions conserved quantities from the old mesh to the new one. This is most
often done advectively (the remesh step is absorbed into the hydrodynamics solve), though in some
instances a direct (geometric) remap is performed.

Here we must also note that ALE schemes typically use grids whose topology is fixed. In the
case of non-simplicial (i.e. hexahedral) grids, this leads to cell faces whose points are not
coplanar. In some instances, cells are defined using curvilinear surfaces, giving a higher-order
scheme (e.g. \citealt{anderson2015}).
Others (e.g. \citealt{garimella2007}) decompose faces into triangles to give a consistent,
polyhedral description of a mesh cell regardless of topology. One exception is the work of
\cite{springel2010}, who solves the hydrodynamic equations on a moving Voronoi mesh, which naturally
gives polyhedral cells. Another is the ``re-ALE'' scheme pioneered by \cite{loubere2010}, in which
the mesh topology is not fixed, but is
``reconnected'' at every timestep to ensure that the mesh remains polyhedral.
The applicability of our method is limited to such polyhedral cells.

As we show in Section \ref{sec:clipping}, a core component of the method presented here is the
clipping operation, in which we construct a convex polyhedron by intersecting a cube with the four
face planes comprising a tetrahedron.
Although we developed this method and optimized our particular implementation
for voxelizing tetrahedra to a grid of
uniform cubes, the clipping operation is general and can be applied to
any convex polyhedron with an arbitrary number of faces. Hence, it may form the basis for such a
direct remap scheme in a hydrodynamics context.
We give a demonstration of such a direct first-order remesh in 2D in Section \ref{sec:2d}.

The problem of intersecting arbitrary polyhedra for direct remapping in hydrodynamics 
has been previously addressed by \cite{grandy1999}, \cite{dukowicz1987}, and \cite{dukowicz1991}. 
The improvement that our method offers over previous work in this area is geometrical robustness. The
aforementioned publications require some auxiliary way of handling geometric degeneracies
(\emph{post-facto} checks on accuracy in the case of \citealt{dukowicz1991}, and \emph{ad-hoc} handling
of all possible degenerate situations in the case of \citealt{grandy1999}). As discussed in Section
\ref{sec:clipping}, our clipping method is automatically robust and thus requires no such checks.

\section{Algorithm} \label{sec:algorithm}

We now describe our voxelization algorithm in detail.
Note that for our application, we have restricted the problem to voxelizing tetrahedra. However, the
concepts can be easily extended into a conservative remesh operation between any meshes, as long as
they consist of convex polyhedra.

The key idea in the voxelization process is to recognize that each intersection between an input
tetrahedron and a voxel is itself a convex polyhedron whose volume and moments can be calculated using
a simplicial decomposition.

This explanation can be made clearer by noting that there are really three types of voxels we must consider:
\begin{enumerate}
	\item Voxels which lie completely inside of the input tetrahedron.
	\item Voxels which lie completely outside of the input tetrahedron.
	\item Voxels which cross the boundary of the input tetrahedron.
\end{enumerate}

Types 1 and 2 are trivial to deal with. Voxels lying completely outside the
tetrahedron can be ignored, and voxels lying completely inside can be integrated over analytically,
since they are axis-aligned cubes. Type 3 is the core of the algorithm because
it requires us to construct the polyhedral domain formed by the intersection of the voxel and the
tetrahedron in a geometrically robust way, a nontrivial operation.

Hence, we break the algorithm into three main parts:

\begin{itemize}
	\item[] \textbf{Searching} is the operation of differentiating between the three voxel types; it amounts to
efficiently finding voxels of type 3.
	\item[] \textbf{Clipping} is the operation of taking a type 3 voxel and constructing the polyhedron resulting from its
intersection with the input tetrahedron.
	\item[] \textbf{Reduction} is the operation of computing the integral of a polynomial density field over voxels of
types 1 and 3. Voxels of type 1 can be reduced trivially, as noted above. Clipped voxels of type 3 must be
decomposed into simplices for computation of the integral. It is this last step that dominates the
reduction operation, so for practical purposes, we use ``reduction'' to mean the combined process of simplicial
decomposition and integration of clipped voxels.
\end{itemize}

\subsection{Searching} \label{sec:search}

We can differentiate between the types of voxels using the orientation of their vertices with
respect to the faces of the tetrahedron. By ``orientation,'' we mean the signed distance from a
vertex at position $\X$ to the face labeled $f$; e.g.

\begin{equation} \label{eq:orient}
d_f = (\X-\X_f) \cdot \bvec{n}_f
\end{equation}

where $\bvec{n}_f$ is the unit normal of the face and $\X_f$ is a point coplanar with the face. Points for
which $d_f \leq 0$ are considered ``behind'' or ``outside of'' $f$, while points for which
$d_f > 0$ are ``inside'' or ``in front of'' $f$.

Voxels with all eight vertices lying outside of the same face of the tetrahedron must be entirely
excluded (type 1), and can be ignored.  Likewise, voxels with all 8 vertices lying inside of all faces of the tetrahedron must be entirely
included (type 2), and can be integrated over easily. Voxels that fall into neither of the above
categories are close to the boundary of the tetrahedron (type 3),
and must be clipped and reduced.

Testing vertices of the grid against the faces of the tetrahedron is a time-consuming operation, especially
if the grid is fine.

In the most na\"{i}ve implementation, we test each grid vertex against all four faces of the
tetrahedron exactly once, storing the results in a buffer. The number of evaluations of
\eqref{eq:orient} in this
brute-force approach scales as $\mathcal{O}(g^3)$, where $g$ is the linear dimension of the grid.

\begin{figure}[h!]
\centering
\begin{subfigure}[b]{0.30\textwidth}
        \includegraphics[width=\textwidth]{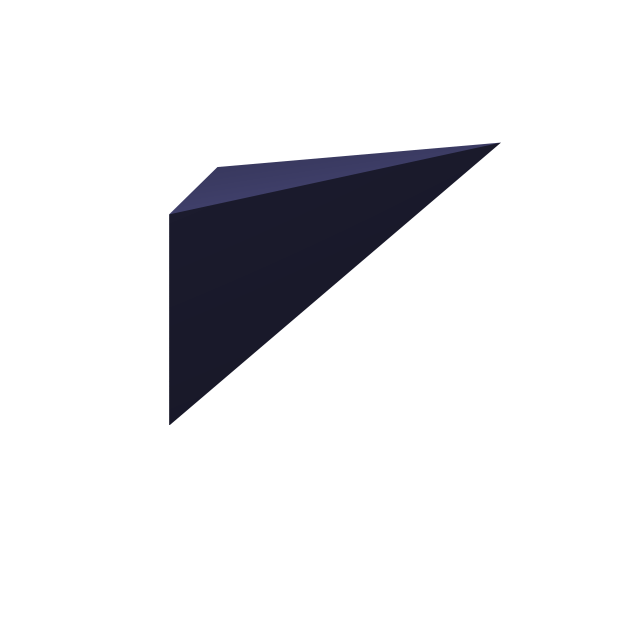}
\end{subfigure}%
\begin{subfigure}[b]{0.30\textwidth}
        \includegraphics[width=\textwidth]{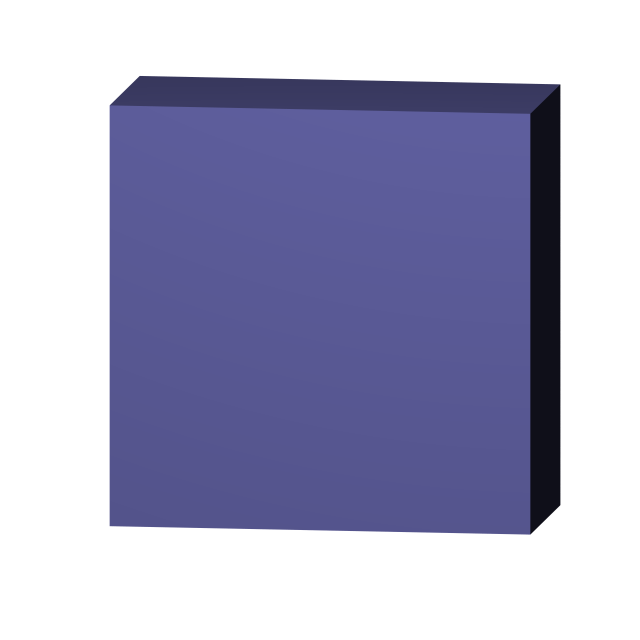}
\end{subfigure}%
\begin{subfigure}[b]{0.30\textwidth}
        \includegraphics[width=\textwidth]{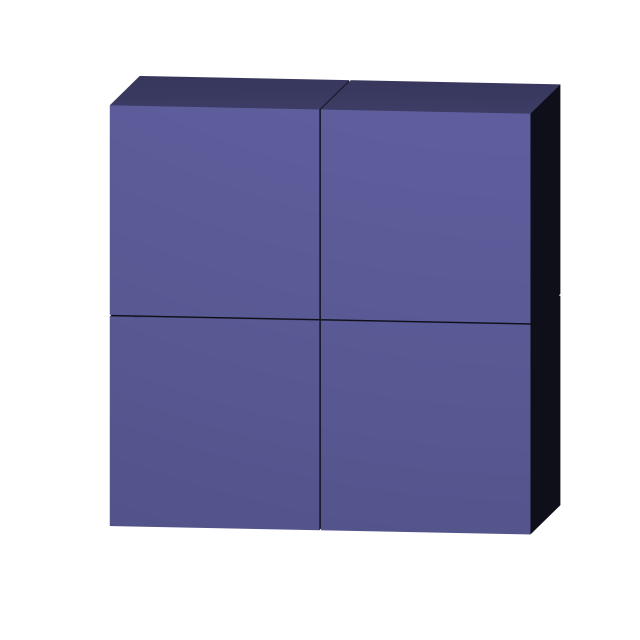}
\end{subfigure}%

\begin{subfigure}[b]{0.30\textwidth}
        \includegraphics[width=\textwidth]{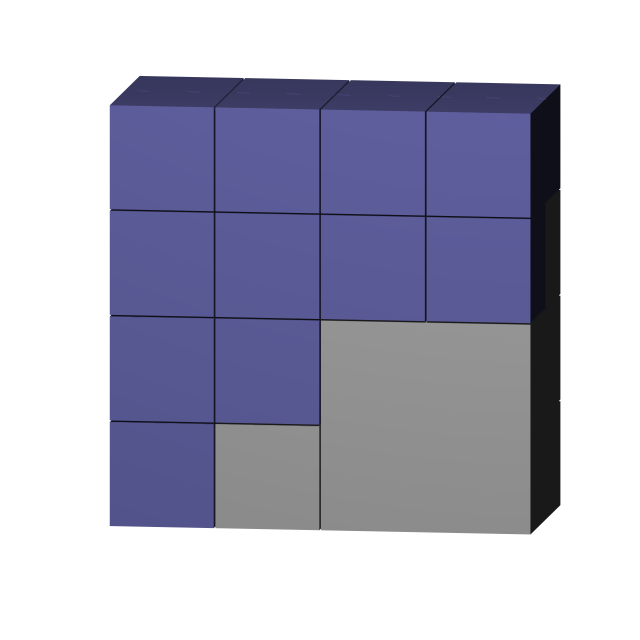}
\end{subfigure}%
\begin{subfigure}[b]{0.30\textwidth}
        \includegraphics[width=\textwidth]{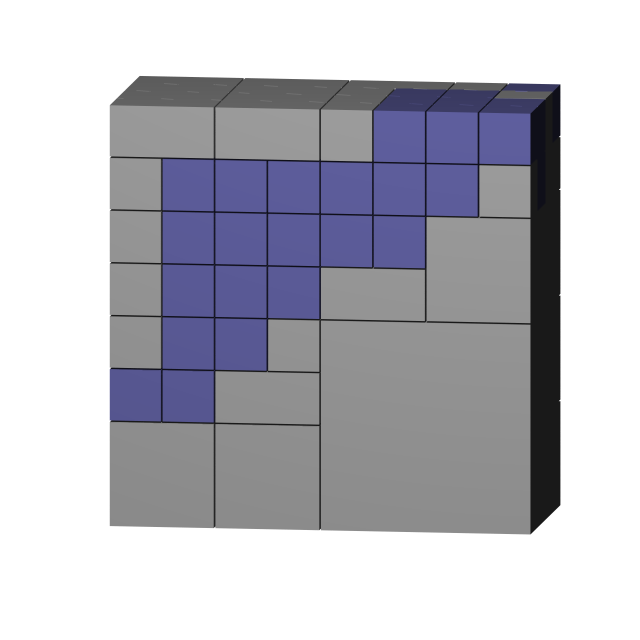}
\end{subfigure}%
\begin{subfigure}[b]{0.30\textwidth}
        \includegraphics[width=\textwidth]{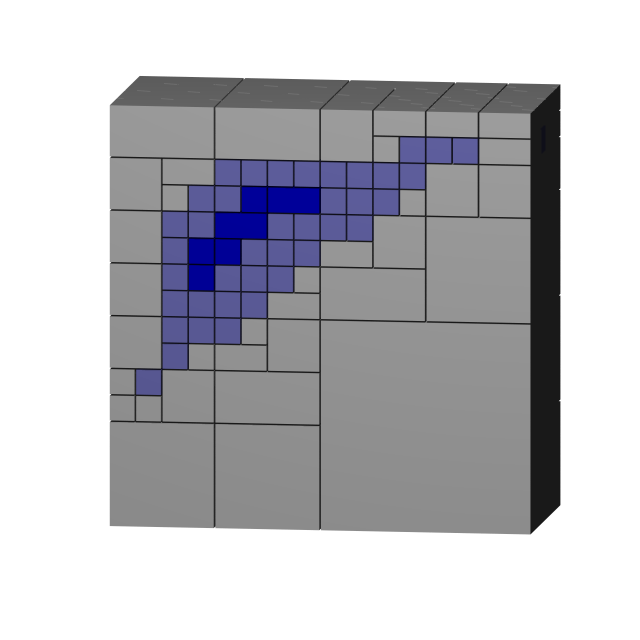}
\end{subfigure}%

\begin{subfigure}[b]{0.30\textwidth}
        \includegraphics[width=\textwidth]{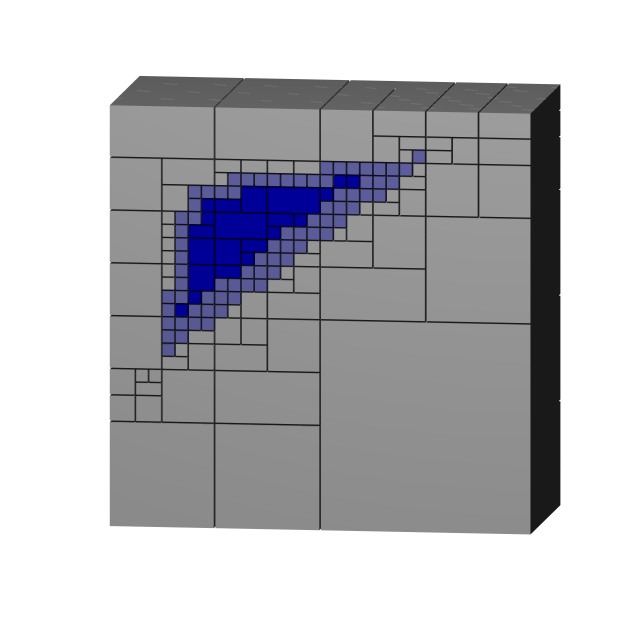}
\end{subfigure}%
\begin{subfigure}[b]{0.30\textwidth}
		\includegraphics[width=\textwidth]{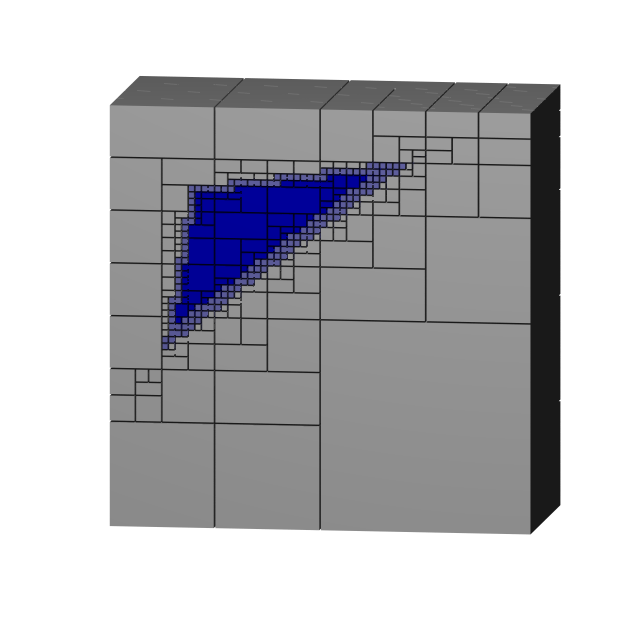}
\end{subfigure}
\begin{subfigure}[b]{0.30\textwidth}
        \includegraphics[width=\textwidth]{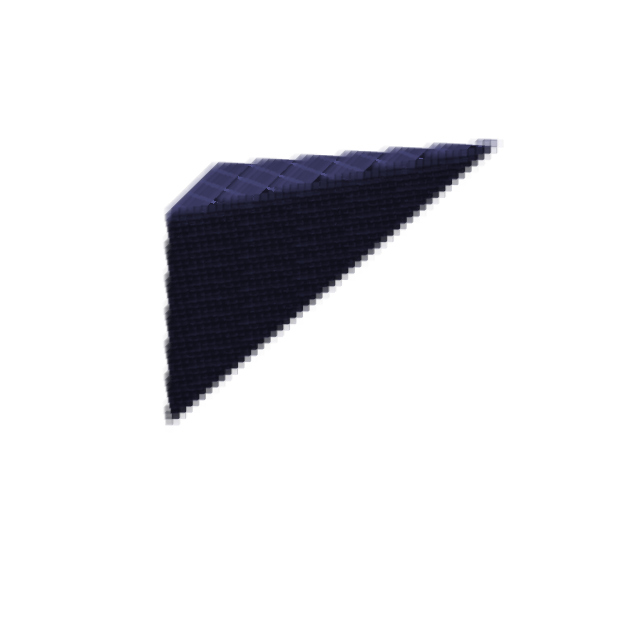}
\end{subfigure}

\caption{A binary space partitioning scheme for minimizing grid vertex orientation checks.
Top left: An input tetrahedron to be voxelized. Top center to bottom center: Recursive
refinement of grid regions. Corners of the grid are checked for inclusion in the tetrahedron.
Regions that lie entirely outside of the
tetrahedron (gray) can be ignored, whereas regions that are entirely included (dark blue)
can be integrated over easily using a for-loop. Regions whose inclusion is ambiguous (light
blue) must lie near the boundary, and are recursively split until they contain a single
voxel. Bottom right: The voxelized tetrahedron.}
\label{fig:bsp}
\end{figure}

It is possible to do better using a binary space partitioning scheme. Instead of checking the
vertices of individual voxels against the faces of the tetrahedron, we begin by checking the corner vertices of
the entire target grid. We then split this box in two across the longest dimension and check the
vertices of those children. This recursive splitting process continues until one of three things
happens:
\begin{enumerate}
	\item All eight vertices of the current box lie outside of the same face of the tetrahedron, so all
		enclosed voxels are completely outside of the tetrahedron. We
		stop recursing and ignore all voxels in the box.
	\item All eight vertices of the current box lie inside of all faces of the tetrahedron, so all
		voxels in the box must be fully contained in the tetrahedron. We
		stop recursing and process all voxels in the box using a for-loop.
	\item The current box contains a single voxel, which must lie on the boundary of the tetrahedron. We stop recursing, clip, and reduce the voxel.
\end{enumerate}
Figure \ref{fig:bsp} gives an illustration of this binary partitioning process.

The number of vertex checks in this binary partitioning approach scales as $\mathcal{O}( g^2 \log g
)$. This is better in principle than the brute-force approach; however, the added overhead of keeping
track of a stack of tree nodes means that practically speaking, the brute-force approach does better
when the target grid is coarse. We discuss this further in the results section.

\subsection{Clipping} \label{sec:clipping}

Any voxels found to lie across the boundary of the tetrahedron must be clipped against the faces of
the tetrahedron, so that the polyhedron resulting from the intersection of the voxel and the
tetrahedron can be known explicitly. By ``clipping'' we mean ``truncation'': we are finding the intersection of
a polyhedron with a half-space (the volume of the original polyhedron that is ``in front of the clip
plane''), throwing away vertices lying outside of the half space (``behind the clip plane''), and
inserting new vertices along edges bisected by the clip plane.

We accomplish this through a directed graph traversal. By Steinitz's theorem (\citealt{steinitz1922}), any convex
polyhedron can be represented as a
three-vertex-connected planar graph (sometimes called a polyhedral graph for this reason) whose vertices
and edges are isomorphic to those of the polyhedron. We use this theorem
to our benefit by representing polyhedra using their graphs, in a modification of the
well-known ``half-edge'' or ``doubly-connected edge list'' representation for polyhedra. This one-skeleton of
the polyhedron provides us with all the necessary information for the geometric operations described
here.
The basic idea of using this representation for robust clipping is reported by \cite{sugihara1994},
though the algorithm is described rather abstractly. We give a concrete implementation which uses a graph traversal over the
polyhedron itself.

The problem of clipping a polyhedron against a plane then reduces to finding the connected
component of its graph whose vertices lie behind the clip plane.

We store polyhedra as triply-linked sets of vertices, where each vertex consists of a coordinate for
its spatial location, three pointers to its neighboring vertices, a byte used to indicate
whether the vertex has been clipped, and four floating-point numbers giving the signed distance to each
clip plane.  Although each vertex this data structure is formally adjacent to three edges, we can
effectively represent an arbitrary number of edges per vertex by employing spatially degenerate vertices
connected by edges of zero length; this feature is essential for geometric robustness.  
Faces are represented implicitly, as loops in the planar graph naturally give us
vertices in the proper order around each face.

The algorithm is as follows. First, the graph of the initial cubical voxel is initialized.
We then traverse the graph using a depth-first search. We begin by finding a vertex behind the clip
plane, simply looping over existing vertices until one is found (this is fast, as the floating-point
operations involved have been previously evaluated). If we cannot find a starting
vertex, we have determined that the entire polyhedron lies in front of the clip plane, and so will
be unaffected.  Otherwise, we start at this vertex and begin traversing the graph. Vertices visited
in the traversal are marked as clipped and ignored hereafter.

Each time a vertex is visited which is in front of the clip plane, we calculate the intersection point
between the clip plane and the edge formed by the current and previous vertices (see Section \ref{sec:waving}
for further explanation). A new vertex is
assigned the correct position, its signed distance to each remaining clip plane is calculated, and
it is inserted into the graph. The previous vertex is marked as having been clipped.

When the traversal ends (all edges behind or crossing the clip plane have been visited), we have in
hand an explicit representation of the clipped voxel, with no need to reconstruct the vertex
ordering. We then repeat the process for each remaining
face of the input tetrahedron.

\begin{figure}[h!]
\centering
\begin{subfigure}[b]{0.24\textwidth}
    \includegraphics[width=\textwidth]{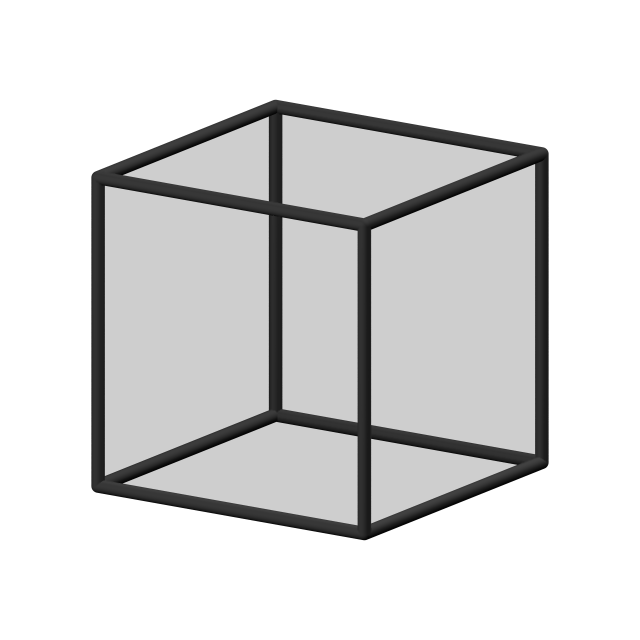}
	%\caption{}
\end{subfigure}%
\begin{subfigure}[b]{0.24\textwidth}
    \includegraphics[width=\textwidth]{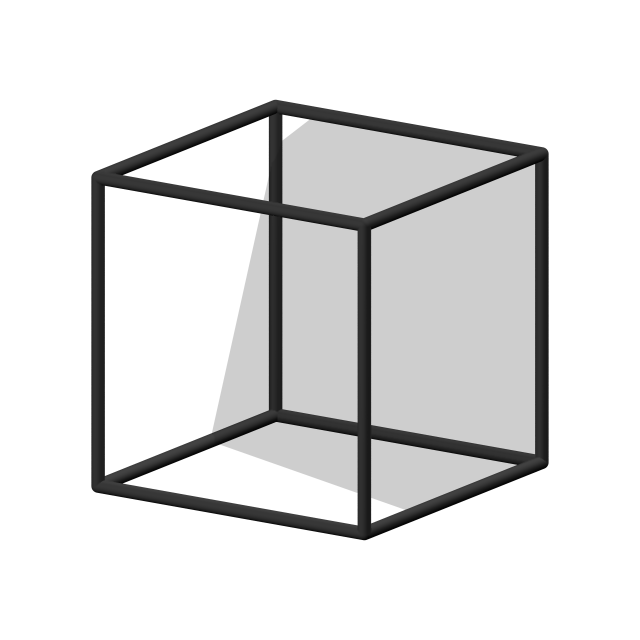}
    %\caption{}
\end{subfigure}%
\begin{subfigure}[b]{0.24\textwidth}
    \includegraphics[width=\textwidth]{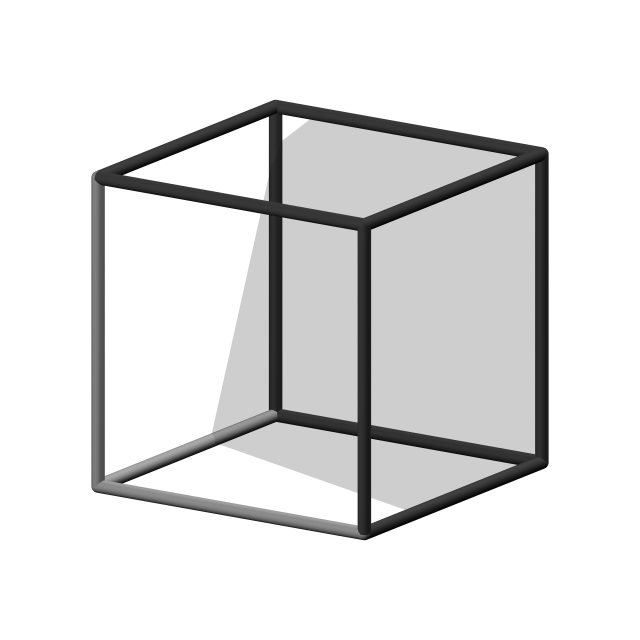}
    %\caption{}
\end{subfigure}%
\begin{subfigure}[b]{0.24\textwidth}
    \includegraphics[width=\textwidth]{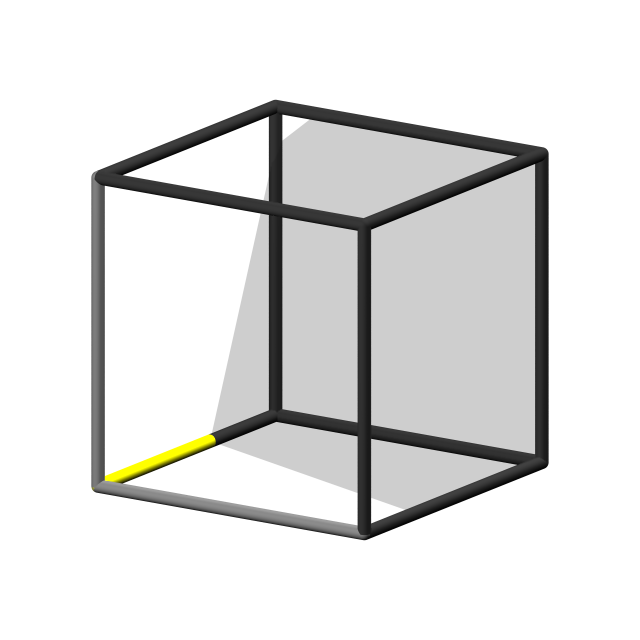}
    %\caption{}
\end{subfigure}%

\begin{subfigure}[b]{0.24\textwidth}
    \includegraphics[width=\textwidth]{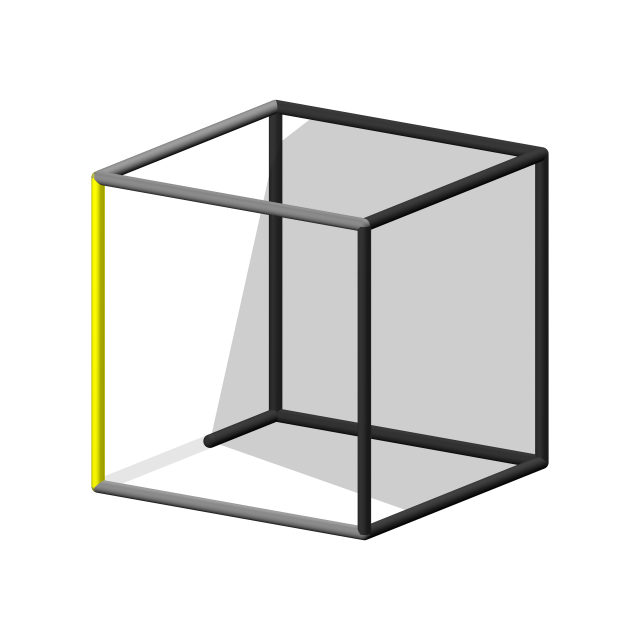}
    %\caption{}
\end{subfigure}%
\begin{subfigure}[b]{0.24\textwidth}
    \includegraphics[width=\textwidth]{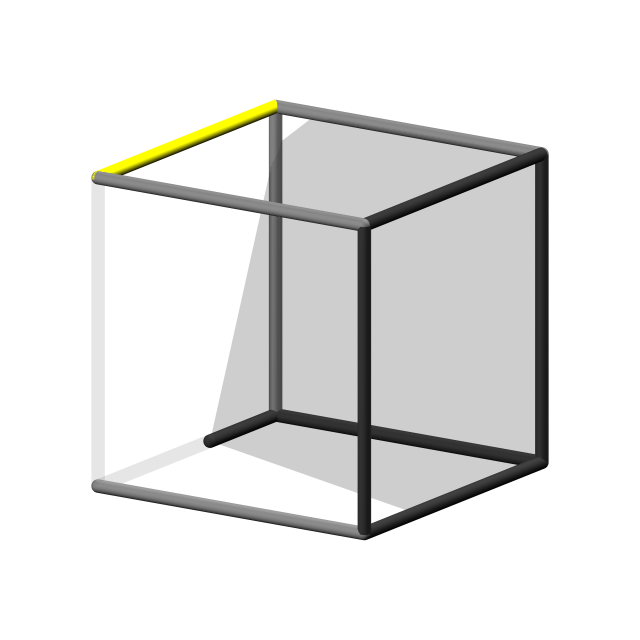}
    %\caption{}
\end{subfigure}%
\begin{subfigure}[b]{0.24\textwidth}
    \includegraphics[width=\textwidth]{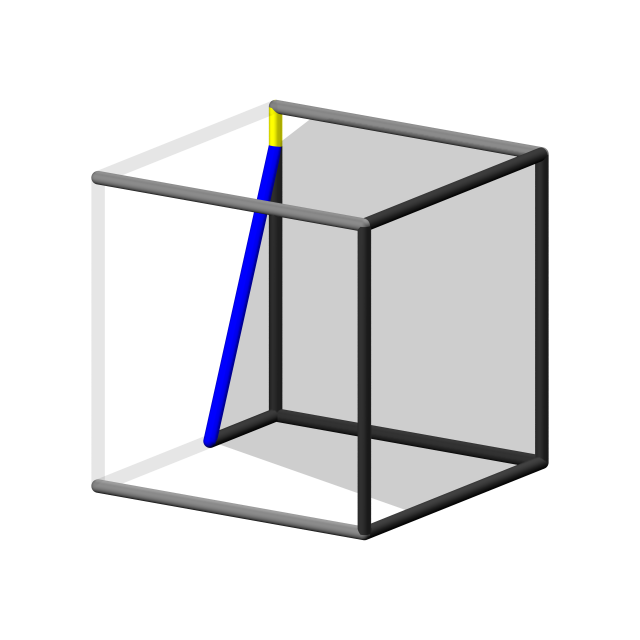}
    %\caption{}
\end{subfigure}%
\begin{subfigure}[b]{0.24\textwidth}
    \includegraphics[width=\textwidth]{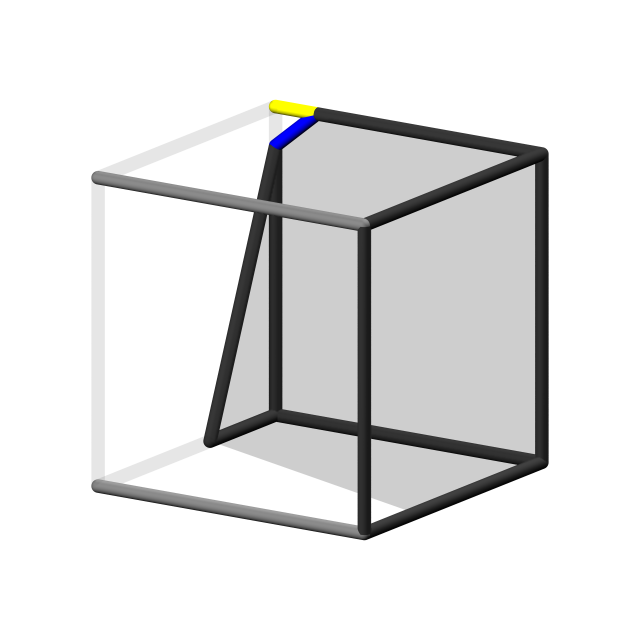}
    %\caption{}
\end{subfigure}%

\begin{subfigure}[b]{0.24\textwidth}
    \includegraphics[width=\textwidth]{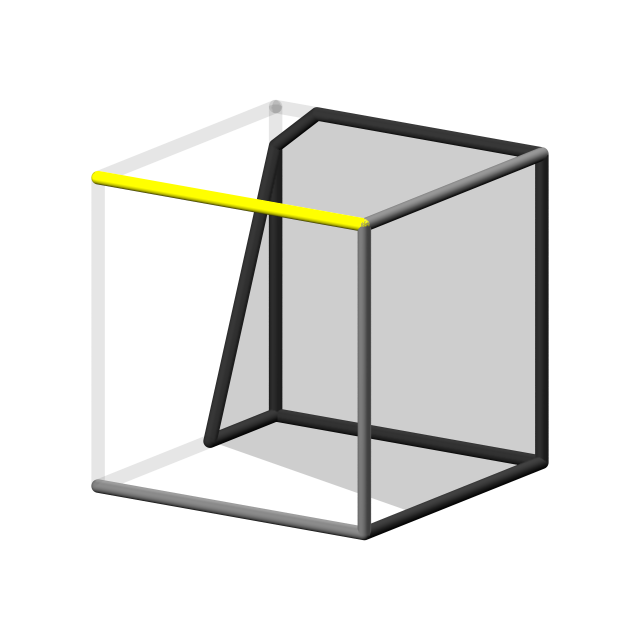}
    %\caption{}
\end{subfigure}%
\begin{subfigure}[b]{0.24\textwidth}
    \includegraphics[width=\textwidth]{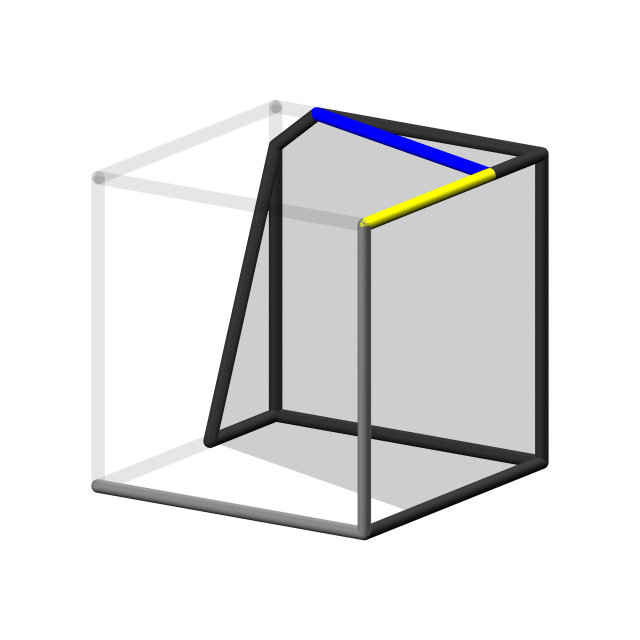}
    %\caption{}
\end{subfigure}%
\begin{subfigure}[b]{0.24\textwidth}
    \includegraphics[width=\textwidth]{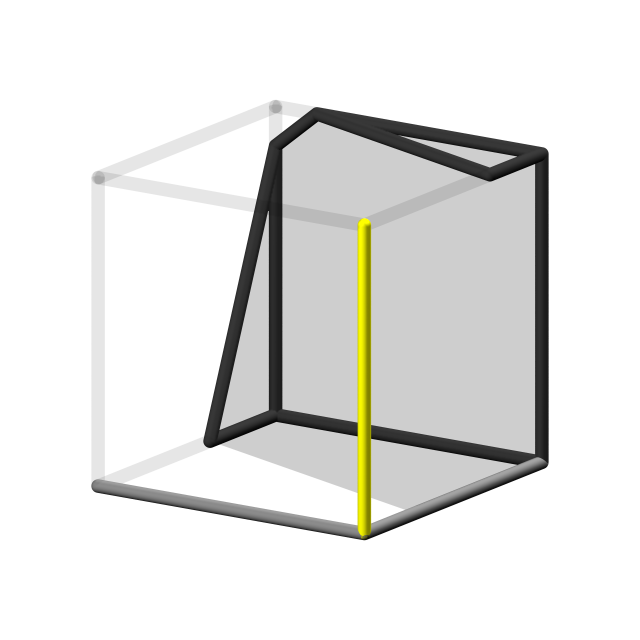}
    %\caption{}
\end{subfigure}%
\begin{subfigure}[b]{0.24\textwidth}
    \includegraphics[width=\textwidth]{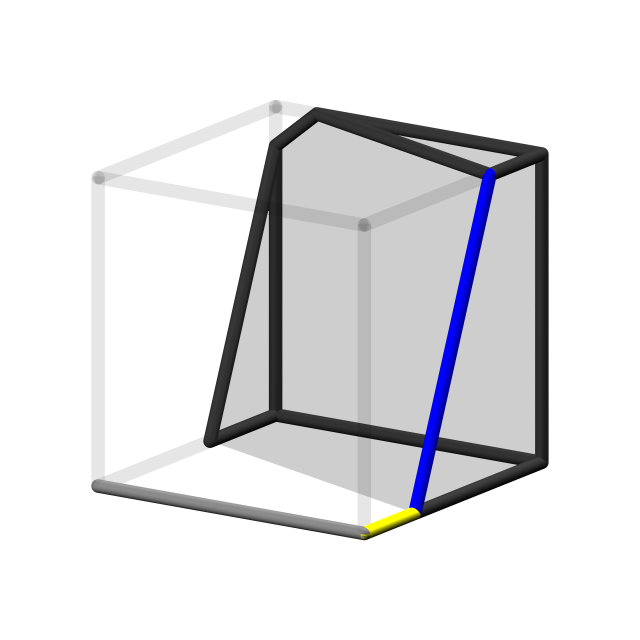}
    %\caption{}
\end{subfigure}%

\begin{subfigure}[b]{0.24\textwidth}
    \includegraphics[width=\textwidth]{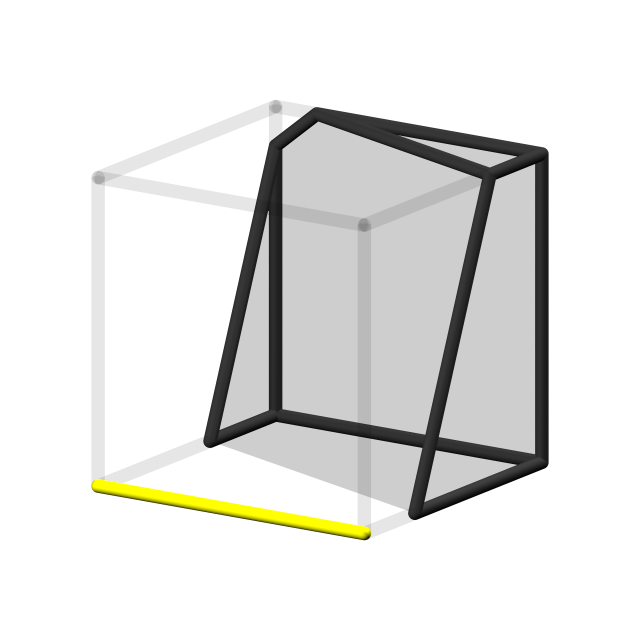}
    %\caption{}
\end{subfigure}%
\begin{subfigure}[b]{0.24\textwidth}
    \includegraphics[width=\textwidth]{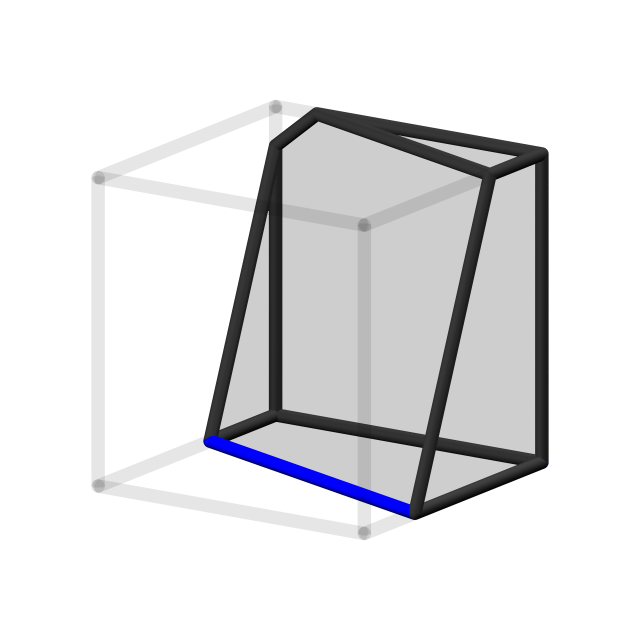}
    %\caption{}
\end{subfigure}%
\begin{subfigure}[b]{0.24\textwidth}
    \includegraphics[width=\textwidth]{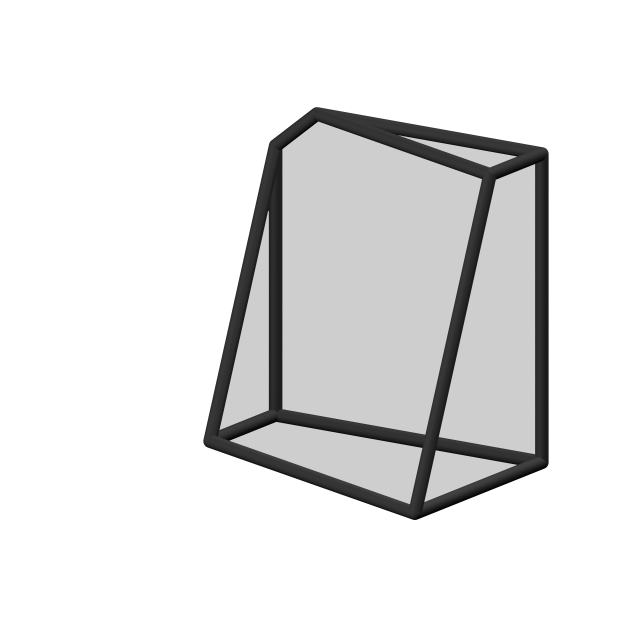}
    %\caption{}
\end{subfigure}%

\caption{Traversal of edges in a voxel for clipping against a plane while automatically
building connectivity of the resulting polyhedron. Top left: The input voxel. Top, second
from left: The shaded region indicates the location of the clip plane. Remaining panels:
The planar graph formed by the vertices and edges of the voxel is traversed. Each time the
working edge (yellow) is bisected by the clip plane, a new edge (blue) is inserted. The
working edge makes its way around the clip plane in an ordered fashion, so that new edges
and vertices are automatically inserted with the appropriate connectivity. Bottom right: The
clipped voxel.}
\label{fig:clip}
\end{figure}

An illustration of the edge traversal process for clipping is given in Figure \ref{fig:clip}.

Approaching the clipping operation in this way has two major benefits:
\begin{enumerate}
	\item We recover new vertices in the correct order (e.g., clockwise around the clip face)
		due to the directed nature of the depth-first graph traversal. This allows us to
		insert new vertices on the fly by implicitly knowing the connectivity to neighboring
		vertices, which saves us from having to reconstruct the polyhedron after each clip
		operation. As we show in Section \ref{sec:results}, this provides substantial performance
		gains over previous methods.
	\item We ensure that the geometrical information in the output is complete and consistent. In
		graph-theoretical terms, we always guarantee preservation of the planar, three-vertex-connected nature of the graph
		while inserting and removing vertices during the clipping process. This makes our algorithm
		robust to degenerate geometry.
\end{enumerate}

\subsection{Reduction} \label{sec:reduction}

Once we have finished clipping a voxel, we are ready to calculate the integral of the input density over the
clipped voxel.  We do this using a simplicial decomposition, e.g. we represent the clipped voxel as
the union of a set of tetrahedra, so that the total integral can be found by summing the integrals over
each tetrahedron.

\begin{figure}[h!]
\centering
\begin{subfigure}[b]{0.32\textwidth}
	\includegraphics[width=\textwidth]{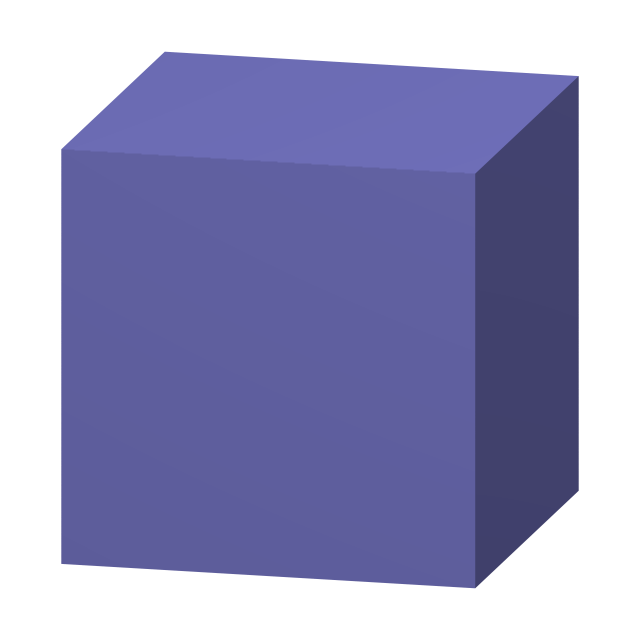}
\end{subfigure}
\begin{subfigure}[b]{0.32\textwidth}
	\includegraphics[width=\textwidth]{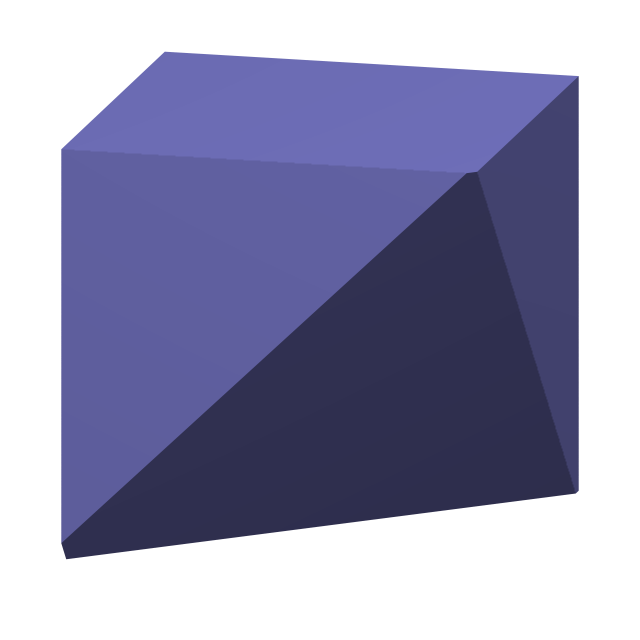}
\end{subfigure}
\begin{subfigure}[b]{0.32\textwidth}
	\includegraphics[width=\textwidth]{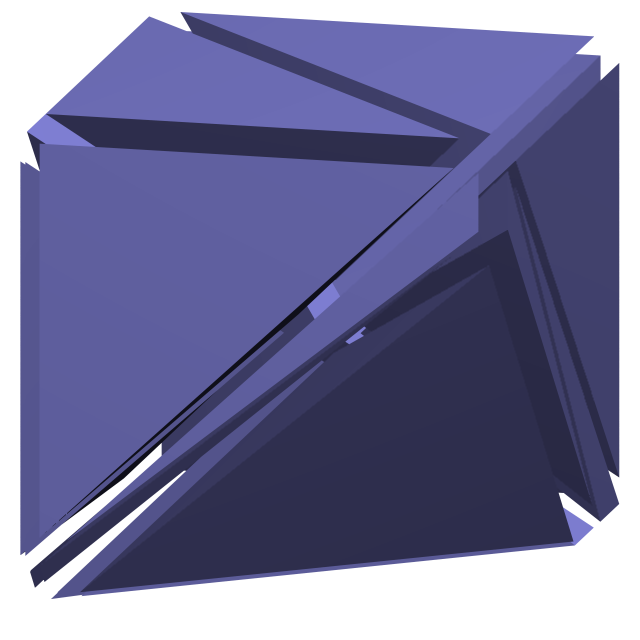}
\end{subfigure}
\caption{An illustrated summary of our method. Left: An initial cubical voxel. Middle: The voxel, clipped
against faces of a tetrahedron, giving a convex polyhedral domain over which we must integrate.
Right: The clipped voxel, decomposed into simplices for integration.}
\label{fig:clip3d}
\end{figure}

The decomposition is also based on a directed graph traversal over the edges. We construct faces
on the fly using the fact that loops in the graph are a natural representation for the faces of the polyhedron.
Because the clipping process is robust, we can assume that vertices on a loop are coplanar.
So, each time we process a new face, we save the first vertex and then loop around the edges. Each
edge, the starting vertex, and the origin form a fan of tetrahedra. It is over
each of these tetrahedra that we integrate before finally summing the results over the entire
decomposition. See Figure \ref{fig:clip3d} for an illustration of the clipping and reduction
process.

One vertex of every tetrahedron in the decomposition is fixed at the origin, to eliminate several floating-point operations.
Due to the linearity of the integrals and the orientation of the tetrahedra, the sum is always correct, even if
the origin lies outside of the original clipped voxel.

Now we discuss the integration itself.  The volume of a tetrahedron is given in terms of the vertex coordinates by the determinant

\begin{equation} \label{eq:tetvol}
V = \frac{1}{6}\begin{vmatrix}
1 & 1 & 1 & 1 \\
x_0 & x_1 & x_2 & x_3 \\
y_0 & y_1 & y_2 & y_3 \\
z_0 & z_1 & z_2 & z_3 \\
\end{vmatrix}
\end{equation}

The integral of a polynomial over a tetrahedral domain $\tetdom$ has been well-known in the finite
element community for some time. \cite{eisenberg1973} give a proof for the following formula:

\begin{equation} \label{eq:baryint}
\int_\tetdom \, \zeta_0^a \, \zeta_1^b \, \zeta_2^c \, \zeta_3^d \, \dtet
= 6\,V\,\frac{ a! \, b! \, c! \, d! }{ (a+b+c+d+3)!}
\end{equation}

where $\zeta_i$ are the barycentric coordinates, $a$, $b$, $c$, and $d$ are integer exponents, and V is
the volume of the tetrahedron (found using \eqref{eq:tetvol}).

Because the integral is given in terms of the barycentric coordinates, we must express it
in terms of Cartesian coordinates. We make use of the fact that
$$
\X = \X_0 \, \zeta_0 + \X_1 \, \zeta_1 + \X_2 \, \zeta_2 + \X_3 \, \zeta_3
$$

where $\X_i$ are the vertex coordinates of the tetrahedron. This allows us to explicitly evaluate an
integral expressed in Cartesian coordinates by substituting the above relation. For example,

\begin{align*}
\int_\tetdom \, x^2 \, \dtet
&= \int_\tetdom \, (x_0 \, \zeta_0 + x_1 \, \zeta_1 + x_2 \, \zeta_2 + x_3 \, \zeta_3)^2 \, \dtet \\
&= \frac{V}{10}(x_0^2+x_1^2+x_2^2+x_3^2+x_0 x_1 + x_0 x_2 + x_0 x_3 + x_1 x_2 + x_1 x_3 + x_2 x_3)
\end{align*}

In the case where we allow $x_0$ to lie at the origin, we can simplify further to

$$
\int_\tetdom \, x^2 \, \dtet
= \frac{V}{10}(x_1^2+x_2^2+x_3^2+ x_1 x_2 + x_1 x_3 + x_2 x_3)
$$

This recipe is the same for all coordinate moments (see Section \ref{sec:moments}).

\section{Practical considerations} \label{sec:practical}

\subsection{Calculation of new vertex locations} \label{sec:waving}

We use a weighted averaging procedure to calculate the positions of new vertices.

Consider two vertices $\X_0$ and $\X_1$ that form the endpoints of an edge that is bisected by a
clip plane. In other words, let $\X_0$ lie in front
of the plane, so that its signed distance to the plane $d_0 > 0$. Let $\X_1$ lie behind
the plane, so that its signed distance to the plane $d_1 \leq 0$. We can calculate the point of
intersection $\X_p$ between the edge and the plane as

\begin{equation}
\X_p = \frac{d_0 \, \X_1 - d_1 \, \X_0}{d_0 - d_1}
\end{equation}

Due to strict use of inequalities in the clipping routine, we guarantee the denominator $d_0 - d_1 > 0$ even in finite precision, so
we are automatically protected from divide-by-zero errors.

The main advantage to using this approach is that it is numerically far more stable than calculating
the intersection point using the plane and the edge in parametric form, especially when $\X_0$,
$\X_1$, or both, lie very near to the plane. Even if $\X_0$ and $\X_1$ are nearly coplanar, the
resulting vertex will lie between the two.

A similar weighted averaging approach is used to calculate the signed distance from the new vertex
$\X_p$ to each of the remaining faces. So, once we find the signed distance to each vertex of a voxel pre-clipping, there is no need to
further keep track of any additional information regarding the faces.

\subsection{Cancellation error} \label{sec:err}

We note that calculating volumes and moments using simplicial
decomposition (see Section \ref{sec:reduction}) can give rise to cancellation errors when evaluated on a computer.

Consider a tetrahedron whose vertices lie in the box
$(x-\Delta x, x-\Delta x, x - \Delta x)$, $(x, x, x)$, where $|\Delta x| < |x|$ (this is a worst-case example in which all
coordinates are larger than the size of the tetrahedron).  The volume calculation $V \sim
\mathcal{O}(\Delta x^3)$ using \eqref{eq:tetvol} involves the subtraction of terms of order $\mathcal{O}(x^3 -
\Delta x^3)$ from terms of order $\mathcal{O}(x^3)$.

We can estimate the error due to cancellation as follows. $b$, the number of significant bits lost
during subtraction, is approximately
\begin{align*}
2^{-b} &\approx 1-\frac{(x^3 - \Delta x^3)}{x^{3}} \\
	&\approx \left(\frac{\Delta x}{x}\right)^{3}
\end{align*}

The fractional error $E \approx 2^{b-p}$ is then dependent on the number of bits $p$ in the significand
($p=53$ in double precision and $p=24$ in single precision for the IEEE
floating-point standard):
\begin{equation} \label{eq:canerror}
E \approx 2^{-p}\, \left(\frac{\Delta x}{x}\right)^{-3}
\end{equation}

This is a toy calculation, and errors will obviously be dependent on the exact geometry in question,
so \eqref{eq:canerror} serves as a rough estimate of how errors should scale with $\Delta x / x$, the size of the
integration domain relative to its absolute coordinate position.

\begin{figure}[h!]
\centering
\includegraphics[width=1.0\textwidth]{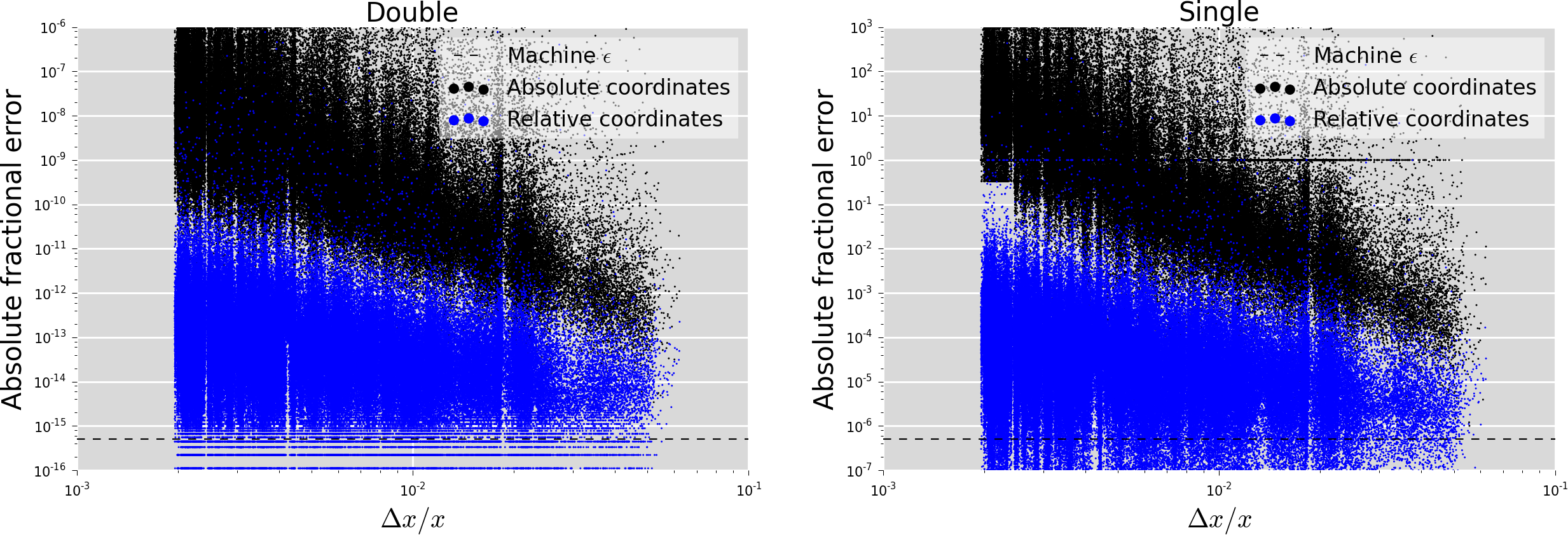}
\caption{Reducing cancellation errors by calculating the volume of each clipped voxel in a
	relative coordinate frame centered on the origin, for double- (left) and single-precision (right)
	calculations. This demonstration was done on a sample of $1.9 \times
10^5$ clipped voxels extracted from 100 pseudo-randomly generated tetrahedra on a $512^3$ grid.
Calculations were validated against the quadruple-precision solution. Errors are
much more well-behaved when we calculate the volume in a relative coordinate frame near the origin.}
\label{fig:errorplot}
\end{figure}

Because of this error scaling, it is beneficial to shift the domain into a local coordinate system near the origin prior
to performing the calculation. In our implementation, we process each voxel with its center lying at the origin.

In shifting the domain, cancellation
errors are still unavoidable. However, doing so changes the scaling of errors in a way that is more
acceptable. Following the same logic as before, subtracting a coordinate offset \emph{prior} to
calculating the volume and moments gives the following error scaling:
\begin{equation} \label{eq:canerror2}
E \approx 2^{-p}\, \left(\frac{\Delta x}{x}\right)^{-1}
\end{equation}

Now, cancellation errors scale linearly with $x / \Delta x$ rather than cubically. So, while some cancellation errors
are unavoidable due to the nature of the problem, we are able to reduce their effect substantially. Figure
\ref{fig:errorplot} shows this scaling for both single- and double-precision calculations. We indeed
see that calculating the volume and moments in a relative coordinate system near the origin is
essential to the overall accuracy of our method.

For higher-order moments, such a coordinate translation gives rise to cross terms that must be taken into account later, but involve only addition and thus do not contribute to cancellation errors. For example, the calculation of the first moment in $x$ over a voxel whose center lies at $x_0$ means first evaluating the moment as though the voxel were centered on the origin, then adding back in a correction for this coordinate offset:

$$
\int_V \, x \, \dV = \int_V \, (x - x_0) \, \dV + x_0 \, V
$$

We note that while this use of a relative coordinate system greatly reduces cancellation errors, any
other numerical error in the calculation of the volume $V$ in the above expression will be scaled by a factor of
$x_0$ and propagated into the final result. The same applies to cross-terms arising in the
evaluation of higher-order moments. This is unavoidable.

We present full results concerning the accuracy of our method in Section \ref{sec:accuracy}.

\subsection{Moments} \label{sec:moments}

Our implementation of this method consists of using the algorithm described above to first find
coordinate moments over each voxel, then taking linear combinations of these moments using
polynomial coefficients. This allows the simultaneous integration of an arbitrary number of scalar
fields (or component-wise integration of vector fields) with no need to re-clip voxels.

For example, consider the integral of the second-order polynomial field
$$
\int_V \, (A x^2 + B y^2 + C z^2 + D x y + E x z + F y z + G x + H y + I z + J) \, \dV
$$
Given the representation of this field in terms of the constant coefficients $A \cdots J$, finding
the integral of the field amounts to finding the integral over each of the coordinate moments
$$
\int_V \, x^2 \, \dV \, \cdots \, \int_V \, xy \, \dV \, \cdots \, \int_V \, x \, \dV \, \cdots \, V
$$

Once all moments for the desired polynomial order are evaluated, any further manipulation (gradient estimation, field normalization, etc.) becomes a linear algebra problem in the space of polynomial coefficients.

\subsection{2D} \label{sec:2d}
All of the concepts presented above extend trivially to two dimensions. The graph traversals
in the clipping and reduction operations
simplify greatly, since the topology of a polygon is a loop.

\begin{figure}[h!]
        \centering
        \includegraphics[width=\textwidth]{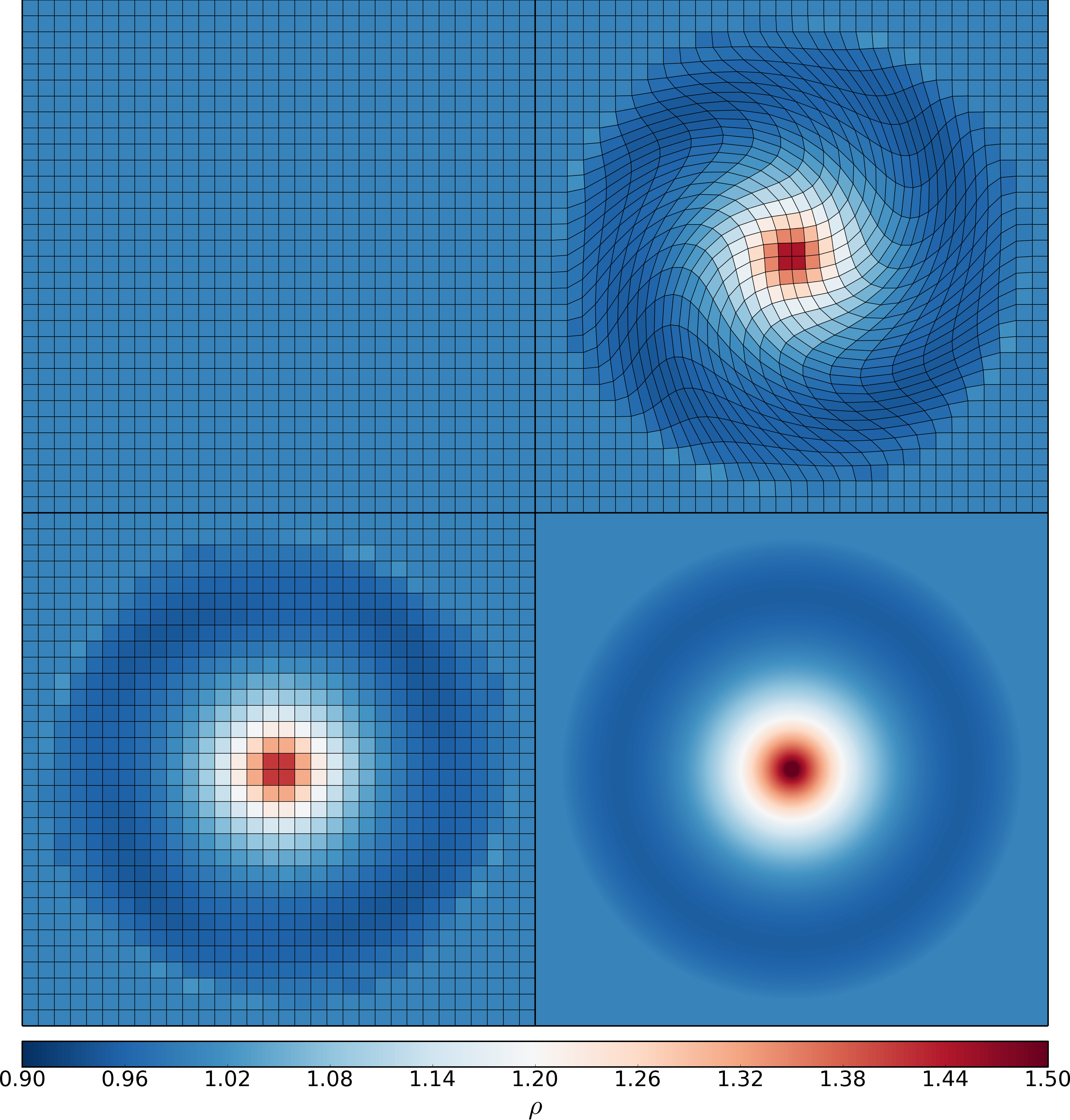}
        \caption{A conservative remesh between quadrilateral grids in 2D. Top left: The initial grid,
		with constant unit density. Top right: The grid, deformed by displacing vertices according
		to the Lagrangian deformation described in Section \ref{sec:2d}. This
	deformation creates a varying density field. Bottom left: The resulting density field, remeshed onto
the original grid. Bottom right: The analytic density for the Lagrangian transformation used.}
\label{fig:remesh2D}
\end{figure}

As a demonstration, we perform a conservative remesh in 2D. We begin with a uniform, Cartesian grid of unit-density quadrilaterals. We then deform the grid by
displacing the gridpoints according to the transformation
(in polar coordinates relative to the center of the grid):
\begin{align*}
	r &\rightarrow r \, S(r/r_0) \\
	\theta &\rightarrow \theta + T(r/r_0)
\end{align*}
where
$$
	S(x) = \left\{
	\begin{array}{ll}
	      1- S_0\,(x-1)^2 &~~~~0 \leq x < 1 \\
	      0 &~~~~x \geq 1
	\end{array}
	\right. \\
$$
and
$$
	T(x) = \left\{
	\begin{array}{ll}
	      T_0 \,(x^2-1)^2 &~~~~0 \leq x < 1 \\
	      0 &~~~~x \geq 1
	\end{array}
	\right. \\
$$

We use the constants $r_0=0.45$ (on a grid of side length 1.0), $S_0=0.2$, and $T_0 = \pi/2$.

The analytic density $\rho(r/r_0)$ resulting from this Lagrangian deformation is given by
$$
\rho(x) = \left\{
	\begin{array}{ll}
		\frac{1}{(S_0\,(x-1)^2-1)\,(S_0\,(3 \, x^2 - 4 \, x + 1) - 1)} &~~~~0 \leq x < 1 \\
	     1 &~~~~x \geq 1
	\end{array}
	\right. \\
$$

We then remesh this deformed grid back onto the original
Cartesian grid. Figure \ref{fig:remesh2D} illustrates this deformation and remesh process.
Total mass is conserved to machine precision. Note that near the boundaries, the old and new meshes are exactly degenerate with (lie exactly on top of)
one another.  This demonstrates the geometrical robustness of our method.

\section{Results and Discussion} \label{sec:results}

In this section, we present the results of numerical tests of our C implementation of this work
(available at \texttt{https://github.com/devonmpowell/r3d}). We first compare results for isolated
clipping and reduction operations against previous work, to give a sense of performance as
a general remesh scheme. We then move on to overall accuracy, timing, and robustness results for voxelizing
tetrahedra.

For fairness in comparisons, we always bring the polyhedron into the appropriate representation for
a method (e.g., planar graph vs. explicit face-vertex connectivity) separately from the operation being timed.

All tests in this section were compiled using \texttt{gcc}, \texttt{g++}, and \texttt{gfortran}
using full optimizations (\texttt{-O3}), and run on a 2.2GHz Intel Xeon E5-2660 processor. Unless
otherwise stated, double-precision arithmetic was used in all computations.

\subsection{Clipping}

We compare our method to the clipping algorithm by \cite{stephenson1975}, as implemented in Fortran
by \cite{lopez2008} as a part of their VOFTools software.  This software also includes
a function for calculating volumes of polyhedra using the expression
$$
V=\frac{1}{6} \, \left [ \sum_{j=1}^{J} \, (\N_j \cdot \X_{j,1}) \, \N_j \cdot \sum_{i=1}^{I_j} \,(\X_{j,i} \times \X_{j,i+1})\right ]
$$
where $\N_j$ is the unit normal of the $j^\textnormal{th}$ face and $\X_{j,i}$ is the
$i^\textnormal{th}$ vertex around face $j$. We compare the speed of this volume
computation as well. A fuller comparison of speed and accuracy for the reduction process, including
computation of all coordinate moments up to second order, is given in the next section.

For this comparison, we clipped a unit cube against a successively increasing number (1, 2, 3, and 4) of
planes. For each test, we timed $10^6$ iterations. The results are summarized in Table
\ref{tab:cmp_clip}. Our calculation of volumes using simplicial decomposition by traversing the planar
graph is slightly slower than the above expression used by \cite{lopez2008}.
However, the speed of our clipping operation is faster by a factor of 3-4. This is due to the fact that we
are able to insert new vertices on the fly with the correct ordering, whereas \cite{lopez2008}
must post-process the clipped polyhedron to bring new vertices into the correct order.

\begin{table}[h!]
\centering
\begin{tabular}{c|cc|cc}
\multicolumn{1}{l}{} & \multicolumn{2}{c}{\textbf{This work}} & \multicolumn{2}{c}{\textbf{\cite{lopez2008}}}\\
	\cline{2-5}
	\multicolumn{1}{l}{\parbox[t]{3em}{\centering Clip\\planes}} &
	\multicolumn{1}{c}{\parbox[t]{6em}{\centering Clip	(ms)}} &
		  \multicolumn{1}{c}{\parbox[t]{6em}{\centering Clip \& \\reduce (ms)}} &
\multicolumn{1}{c}{\parbox[t]{6em}{\centering Clip (ms)}} &
			\multicolumn{1}{c}{\parbox[t]{6em}{\centering Clip \& \\reduce (ms)}}  \\ \hline \hline
0 &   - & 280 & - & 180 \\
1 &	220 & 580 & 870 & 1080 \\
2 &	370 & 790 & 1360 & 1620 \\
3 &	560 & 1050 & 1920 & 2200 \\
4 &	760 & 1320 & 2550 & 2880 \\
\end{tabular}
\caption{Comparison of timing with the clipping and volume implementations of \cite{lopez2008}. A
cube was clipped against different numbers of planes. Times are given in milliseconds per $10^6$ trials.
Although the volume computation is slower in our implementation due to overhead in the graph
traversal, our representation of the polyhedron as a planar graph gives an overall speed-up, as it
automatically inserts new vertices in the proper order.}
\label{tab:cmp_clip}
\end{table}

\subsection{Reduction} \label{sec:reduc_results}

We now show a comparison of our reduction process with three other methods. The first
is a control consisting of the same simplicial decomposition scheme used here, but implemented using
for-loops rather than the planar graph traversal (the decomposition scheme is the same,
though ordering of operations may vary). The second is the volume computation given by
\cite{lopez2008}. The third is the volume and moments computation using dimensionality reduction, by
\cite{mirtich1996}.

We computed volumes and moments for three polyhedra of varying complexity: a tetrahedron, a cube, and
a dodecahedron, all scaled and translated to lie in the unit cube in the first octant. For each test, we ran $10^6$ iterations. Errors were calculated relative to the solution of \cite{mirtich1996}.
We show results for both timing and accuracy in Table \ref{tab:cmp_reduce}.

\begin{table}[h!]
\centering
\begin{tabular}{l|ccc|ccc}
\multicolumn{1}{l}{} & \multicolumn{3}{c}{\textbf{This work}} &
	\multicolumn{3}{c}{\textbf{Decomposition, for-loop}}\\
	\cline{2-7}
	\multicolumn{1}{l}{} &
	\multicolumn{1}{c}{\parbox[t]{3.5em}{\centering Volume (ms)}} &
		  \multicolumn{1}{c}{\parbox[t]{3.5em}{\centering Moments (ms)}} &
		  \multicolumn{1}{c}{\parbox[t]{3.5em}{\centering Error}} &
	\multicolumn{1}{c}{\parbox[t]{3.5em}{\centering Volume (ms)}} &
		  \multicolumn{1}{c}{\parbox[t]{3.5em}{\centering Moments (ms)}} &
		  \multicolumn{1}{c}{\parbox[t]{3.5em}{\centering Error}} \\ \hline \hline
Tetra.  & 150 & 290 & $7.2\times 10^{-16}$ & 60 & 220 & $7.2\times 10^{-16}$ \\
Cube    & 280 & 670 & $1.7\times 10^{-16}$ & 170 & 610 & $1.7\times 10^{-16}$ \\
Dodec.  & 690 & 1850 & $9.2\times 10^{-16}$ & 500 & 1820 & $7.3\times 10^{-16}$ \\
\end{tabular}
\newline
\vspace*{1em}
\newline
\begin{tabular}{l|ccc|ccc}
\multicolumn{1}{l}{} & \multicolumn{3}{c}{\textbf{\cite{lopez2008}}} &
	\multicolumn{3}{c}{\textbf{\cite{mirtich1996}}}\\
	\cline{2-7}
	\multicolumn{1}{l}{} &
	\multicolumn{1}{c}{\parbox[t]{3.5em}{\centering Volume (ms)}} &
		  \multicolumn{1}{c}{\parbox[t]{3.5em}{\centering Moments (ms)}} &
		  \multicolumn{1}{c}{\parbox[t]{3.5em}{\centering Error}} &
	\multicolumn{1}{c}{\parbox[t]{4.5em}{\centering Volume (ms)}} &
		  \multicolumn{1}{c}{\parbox[t]{3.5em}{\centering Moments (ms)}} &
		  \multicolumn{1}{c}{\parbox[t]{4.5em}{\centering Error}} \\ \hline \hline
Tetra.  & 100 & - & $3.3\times 10^{-16}$ & - & 1050 & 0.0 \\
Cube    & 180 & - & 0.0 & - & 1760 & 0.0 \\
Dodec.  & 400 & - & 0.0 & - & 3980 & 0.0 \\
\end{tabular}
\caption{Comparison of timing and accuracy between reduction methods. Times are given in milliseconds per
	$10^6$ trials. The error quoted is the maximum absolute fractional error from all 10
coordinate moments, except for \cite{lopez2008}, where only volume information is available.
We see that reduction using our graph traversal is slightly slower, though this becomes
less significant when higher-order moments are calculated.}
\label{tab:cmp_reduce}
\end{table}

Our reduction using a graph traversal carries a slight overhead compared with the
for-loop implementation and \cite{lopez2008}, though this overhead becomes much less significant
when higher-order moments are calculated, due to the dominance of floating-point operations in the
computation. All calculations agree to within machine precision.

\subsection{Full voxelization results} \label{sec:accuracy}

To test the overall accuracy of our method in terms of conservation, we voxelized $10^5$
tetrahedra of unit density whose vertices were randomly chosen to lie in the unit cube onto a
$128^3$ grid. This gave a volume of $1.3 \times 10^{-2}$, $2.3 \times 10^4$ interior (type 1) voxels, and
$1.5 \times 10^4$ boundary (type 3) voxels requiring clipping, per tetrahedron on average.
We compare the rms and maximum fractional errors between the moment integrals of the input tetrahedron
and the sum over all voxels in the output. The results are summarized in Table
\ref{tab:rand_nondegen}.

\begin{table}[h!]
\centering
\begin{tabular}{l|cc|cc}
\multicolumn{1}{l}{} & \multicolumn{2}{c}{\textbf{Double}} & \multicolumn{2}{c}{\textbf{Single}} \\
	\cline{2-5}
\multicolumn{1}{l}{} & \multicolumn{1}{c}{rms} & \multicolumn{1}{c}{max} & \multicolumn{1}{c}{rms} &
	\multicolumn{1}{c}{max}  \\ \hline \hline
Constant  & $1.7\times 10^{-12}$ & $5.2\times 10^{-10}$ & $5.6\times 10^{-4}$ & $1.4\times 10^{-1}$ \\
Linear    & $1.6\times 10^{-12}$ & $5.4\times 10^{-10}$ & $5.4\times 10^{-4}$ & $1.7\times 10^{-1}$ \\
Quadratic & $1.6\times 10^{-12}$ & $5.7\times 10^{-10}$ & $5.3\times 10^{-4}$ & $1.9\times 10^{-1}$ \\
\end{tabular}
\caption{Accuracy (conservativeness) of our voxelization method in double- and single-precision, for
	a sample of $10^5$ randomly-generated tetrahedra on a $128^3$ grid.
Here we quote the maximum fractional error between the pre- and post- voxelization moments. Errors for
each polynomial order are taken from all moments in that order (e.g. ``linear'' includes $x$, $y$,
and $z$ moments).}
\label{tab:rand_nondegen}
\end{table}

\comments{

Double:
-6.93889390401e-18
1 + 2.22044604925e-16
5.2e-10 & 1.7e-12 \\
5.4e-10 & 1.6e-12 \\
5.7e-10 & 1.6e-12 \\

Single:
-4.96676966222e-09
1 + 1.19209289551e-07
1.4e-01 & 5.6e-04 \\
1.7e-01 & 5.4e-04 \\
1.9e-01 & 5.3e-04 \\

}

We see that, although quite accurate, we are still a few orders of magnitude in accuracy away from
machine precision, even though the reduction process itself is accurate to machine precision (see
Section \ref{sec:reduc_results}). This is due to the unavoidable presence of cancellation errors
that arise in shifting each
voxel to a local coordinate frame prior to integration. See Section \ref{sec:err} for a full
discussion of these errors.

To demonstrate the geometrical robustness of our method, we repeated the same test as before; however, this time we
randomly generated the tetrahedron vertices at integer multiples of the grid spacing. This ensures
that we have exactly degenerate geometry. The results are shown in Table \ref{tab:rand_degen}.

\begin{table}[h!]
\centering
\begin{tabular}{l|cc|cc}
\multicolumn{1}{l}{} & \multicolumn{2}{c}{\textbf{Double}} & \multicolumn{2}{c}{\textbf{Single}}\\
	\cline{2-5}
\multicolumn{1}{l}{} & \multicolumn{1}{c}{rms} & \multicolumn{1}{c}{max} & \multicolumn{1}{c}{rms} &
	\multicolumn{1}{c}{max}  \\ \hline \hline
Constant  & $5.6\times 10^{-14}$ & $7.2\times 10^{-14}$ & $3.3\times 10^{-5}$ & $6.0\times 10^{-3}$ \\
Linear    & $5.8\times 10^{-14}$ & $7.5\times 10^{-14}$ & $3.4\times 10^{-5}$ & $6.7\times 10^{-3}$ \\
Quadratic & $6.1\times 10^{-14}$ & $8.1\times 10^{-14}$ & $3.5\times 10^{-5}$ & $7.7\times 10^{-3}$ \\
\end{tabular}
\caption{The same accuracy test as Table \ref{tab:rand_nondegen}, but for tetrahedra whose vertices
were chosen to be exactly degenerate with the grid. Our method is immune to such geometrical
degeneracies, as demonstrated by the high accuracy of the computations.}
\label{tab:rand_degen}
\end{table}

\comments{
Double:
-2.31296463464e-17
1 + 2.22044604925e-16
7.2e-12 & 5.6e-14 \\
7.5e-12 & 5.8e-14 \\
8.1e-12 & 6.1e-14 \\

Single:
-1.24175771887e-08
1 + 2.38418579102e-07
6.0e-03 & 3.3e-05 \\
6.7e-03 & 3.4e-05 \\
7.7e-03 & 3.5e-05 \\

}

The results indicate that the degenerate case is actually \emph{more} accurate than the general case. This is
due to the fact that we ran this test on a $128^3$ grid, a power of two, to ensure exact geometric degeneracies
in the binary representation. A side effect is that the computations themselves are
more accurate, since they take place in a representation that favors binary fractions of the grid spacing.
So, the numbers in Table \ref{tab:rand_degen} should not be taken as a statement about the general
accuracy of the code (for that, refer to Table \ref{tab:rand_nondegen}), but as a statement of the
geometric robustness of our method.

\begin{figure}[h!]
\centering
\includegraphics[width=1.0\textwidth]{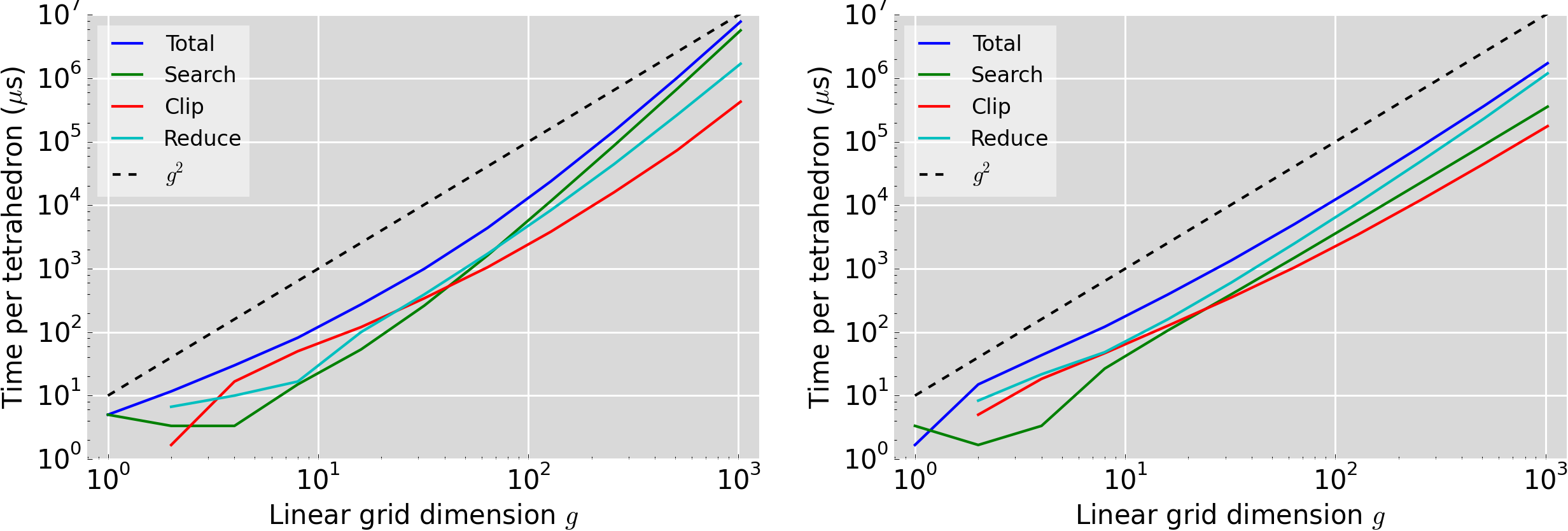}
\caption{Timing of the operations involved in our voxelization method. In both cases, the clip and
	reduce operations scale roughly quadratically in the grid dimension (e.g. linearly with surface
	area), as expected. Left: Using a full grid buffer to search for voxels that need
	further processing makes the search operation scale as the
cube of the grid dimension. Right: Spatial tree for voxel searching. All operations scale
quadratically in the grid dimension (the search operation contains a logarithmic component, but the
quadratic term dominates). The reduction operation also does slightly better, since large
blocks of fully included voxels can be processed together.}
\label{fig:timing_summary}
\end{figure}

As a final test, we check the performance scaling of the voxelization routine with grid resolution. We repeat
the same test as before ($10^5$ tetrahedra with randomly generated, though nondegenerate, vertices),
voxelizing them onto grids of increasing linear resolution $g$, up to $1024^3$. We do this for both
search methods described in Section \ref{sec:search} (brute-force and binary space partitioning).
The results are summarized in Figure \ref{fig:timing_summary}.

We see that, in general, the performance scales as $g^2$, or equivalently, as the effective surface
area of the tetrahedra in units of squared grid spacing. This is consistent with our expectations,
since clipping and reduction are the most costly operations involved, and they take place only on
the boundaries of tetrahedra.  An exception to this scaling arises in the brute-force search method,
which checks every grid point against each tetrahedron face, and so scales as $g^3$ when the grid
becomes sufficiently fine. This is in
contrast to the binary space partitioning search, which scales as $g^2 \log g$. However, in
practical terms, the brute-force method actually performs better for coarser grids due to the added
overhead of the binary search. Binary partitioning only overtakes brute-force at a grid resolution
of $128^3$, corresponding to an average tetrahedron volume of $\sim3 \times 10^4$ voxels.

\section{Conclusion}

We describe a general remeshing method, in that we present an approach to robustly intersecting two convex
polyhedra and computing a polynomial integral over the resulting intersection domain. 

Such an operation is useful for computational physics in several areas. These include ALE and re-ALE
hydrodynamics, in which fluid quantities must be transferred between meshes in a geometrically
precise way, sometimes with higher-order polynomial interpolation 
(e.g. \citealt{Donea2004}, \citealt{loubere2010}, \citealt{dukowicz1987},
and \citealt{dukowicz1991}). Interface
reconstruction and volume-of-fluid methods (\citealt{hirt1981}, \citealt{renardy2001},
\citealt{lopez2008}) also rely on such a geometric intersection followed by an integral. 
Computing exact integrals over the intersection between two polyhedra is also
useful in computer graphics and visualization, where the exact computation of convolution integrals
is of interest (e.g. \citealt{catmull1978}, \citealt{duff1989}, \citealt{auzinger2012}, and
\citealt{auzinger2013}). We focus on yet another application, the exact mass-conservative
voxelization of tetrahedra for the simulation and analysis of cosmological $N$-body systems using
the approach of \cite{AHK2012}. This interpretation of the N-body problem has proven quite useful in
recent work (\citealt{Kaehler2012}, \citealt{hahn2013}, \citealt{angulo2014}, \citealt{hahn2014}).

This general problem of computing exact intersection volumes between polyhedra and integrating
over those volumes has been studied in detail by \cite{dukowicz1987}, \cite{dukowicz1991}, and
\cite{grandy1999}. Additionally, \cite{lopez2008} give an implementation of the basic clipping
operation of \cite{stephenson1975}.  A common issue that these implementations must deal with is 
how to handle geometric degeneracies in the input. For example, \citealt{dukowicz1991} impose
\emph{post-facto} checks on accuracy, while \citealt{grandy1999} employs \emph{ad-hoc} handling
of all possible degenerate situations.

The main contribution of this paper is to put forth a unified framework for the problem of
intersecting convex polyhedra in a geometrically robust way, and subsequently computing an integral
over the resulting domain. The specific case on which we focus is the physically conservative
voxelization of tetrahedra with polynomial densities.
We present a C implementation as well.

Our algorithm for intersecting two convex
polyhedra by successively clipping one against the faces of the other is based on the ideas of
\cite{sugihara1994}, who describes
in abstract terms how the planar graph representation of a polyhedron can be used in a geometrically
robust clipping algorithm by guaranteeing the topological validity of the output. 
Our implementation is based on a depth-first graph traversal, which ensures that it is automatically geometrically
robust, with no need for auxiliary checks or high precision arithmetic.
We couple the clipping algorithm to an integration routine on the same planar graph representation.
As a result, we are able to store polyhedra using only vertex locations and neighbors, with no
need for face normals. 

We address practical issues including numerical stability of
geometric calculations, management of cancellations errors, and extension to two dimensions.
In a comparison to the implementation of \cite{lopez2008},
we show that our clipping operation is faster by a factor of 3-4, with an overall
speed-up by a factor of 2. This is due to the algorithm's ability to insert new vertices in the correct order on the fly,
with no need to reorder them post-clipping. Our implementation conserves the integral between the
input and output meshes to high precision. 

Our C code (available at \texttt{https://github.com/devonmpowell/r3d}) is intended to be a simple
tool for carrying out fast, accurate, and robust geometrical
calculations on the convex polyhedral mesh elements often used in computational physics. 

\section*{Acknowledgements}

We are grateful to T. Sousby and O. Hahn for useful discussions on
this topic. 
T.A. also is grateful Pat Hanrahan for suggesting some classic references and encouragement.
We are indebted to R. Kaehler from whom we have learned a great deal about algorithms,
rasterization, and GPUs.

D. Powell was supported in this work by the Fletcher Jones Foundation Stanford Graduate Fellowship.
This work was also supported in part by the U.S. Department of Energy contract to SLAC no. DE-AC02-76SF00515.

\newpage

\bibliography{main}

\end{document}